\documentclass[11pt,a4paper]{article}

\usepackage[tbtags]{amsmath}

\usepackage[DIV=12]{typearea}
\usepackage{lineno}  

\usepackage[oldstylenums]{kpfonts}

\usepackage[english]{babel}

\usepackage{scrextend} 

\usepackage{setspace}
\usepackage{physics}

\usepackage{url} 
\usepackage[pagebackref]{hyperref} 
\hypersetup{
	pdftitle={Discrete-time maximally superintegrable systems and deformed symmetry algebras: the Calogero--Moser case},
    pdfauthor={Pavel Drozdov, Giorgio Gubbiotti, Danilo Latini},
    colorlinks,
    citecolor=brightviolet,
    filecolor=black,
    linkcolor=brightviolet,
    urlcolor=brightviolet,
    final=true
} 
\usepackage[dvipsnames]{xcolor} 
\definecolor{my-background}{RGB}{255, 242, 204}
\definecolor{brightviolet}{RGB}{148,0,211}  

\usepackage{blkarray}

\usepackage{microtype}

\allowdisplaybreaks  

\usepackage{mathbbol}
\usepackage{braket}
\usepackage{tensor}

\usepackage[tiny]{titlesec}
\usepackage{titletoc}  
\titleformat{\section}
  {\normalfont \bfseries \scshape }{\thesection. }{0.1em}{}

\titlecontents{section}[0pt]{\addvspace{1ex}}{\thecontentslabel. \normalfont \bfseries \scshape}
{\Large\bfseries}{\hspace{1em plus 1fill}\large\bfseries \contentspage}

\titleformat{\subsection}
  {\normalfont \bfseries}{\thesubsection. }{0.1em}{}
  \titleformat{\subsubsection}
  {\normalfont \small \bfseries}{\thesubsubsection. }{0.1em}{}
  
  \renewcommand{\thesection}{\Roman{section}}

\newcommand{\N}{\mathbb{N}}
\newcommand{\R}{\mathbb{R}}
\newcommand{\Z}{\mathbb{Z}}

\newcommand{\1}{\mathbb{1}}

\DeclareMathOperator{\diff}{d\!} 

\newcommand{\mR}{\mathcal{R}}

\newcommand{\su}{\mathfrak{su}}

\newcommand{\gl}{\mathfrak{gl}}

\DeclareMathOperator{\diag}{diag}

\DeclareMathOperator{\Ord}{Ord}

\renewcommand{\imath}{\mathrm{i}}

\renewcommand{\bar}{\overline}

\newcommand{\K}{\widetilde{K}}

\newcommand{\G}{\widetilde{G}}

\newcommand{\M}{\widetilde{M}}

\usepackage{bbold}

\newcommand{\hyphen}{\hbox{--}}
\newcommand{\ie}{\textit{i.e.}}
\newcommand{\mm}{\textit{mutatis mutandis}} 
\newcommand{\eg}{\textit{e.g.}}
  
\newcommand{\cf}{\textit{cf.}}
\newcommand{\vv}{\textit{vice versa}}

\usepackage{enumerate}
\usepackage[shortlabels]{enumitem} 
\setlist[enumerate,1]{label={(\roman*)}}

\usepackage[normalem]{ulem}  

\usepackage{amsthm}
\usepackage{thmtools}
\usepackage[]{cleveref} 

\newtheorem{theorem}{Theorem}[section]
\newtheorem*{theorem*}{Theorem}

\newtheorem{corollary}{Corollary}[theorem]

\newtheorem{prop}[theorem]{Proposition}
\theoremstyle{definition}

\newtheorem*{newton}{Newton's identity}

\theoremstyle{remark}
\newtheorem{remark}[theorem]{Remark}
\newtheorem{example}[theorem]{Example}

\makeatletter
\renewcommand{\fnum@figure}{Fig.\,\thefigure}

\makeatletter
\renewenvironment{proof}[1][\relax]{\par
  \pushQED{\qed}%
  \normalfont \topsep6\p@\@plus6\p@\relax
  \trivlist
  \item[\hskip\labelsep\itshape
    \ifx#1\relax \proofname\else\proofname{} of 
    #1\fi\@addpunct{.}]\ignorespaces
}{%
  \popQED\endtrivlist\@endpefalse
}
\makeatother

\usepackage[auth-sc,affil-it]{authblk}

\title{Discrete-time maximally superintegrable systems and deformed symmetry algebras: the Calogero--Moser case}

\author[1]{Pavel Drozdov} 
\author[2]{Giorgio Gubbiotti}
\author[2]{Danilo Latini} 

\makeatletter
\renewcommand\AB@affilsepx{\\\vspace{0.5em}}
\makeatother

\affil[1]{Dipartimento di Scienze Matematiche, Informatiche e Fisiche, 
  Universit\`a degli Studi di Udine, 33100 Udine, Italy   \&  INFN Sezione di Trieste, 34127 Trieste, Italy  } 
\affil[2]{Dipartimento di Matematica ``Federigo Enriques'',  
Universit\`a degli Studi di Milano \& INFN Sezione di Milano, 20133 Milan, Italy} 
\bgroup

\affil[ ]{\footnotesize \ttfamily \emph{e-mails:}
     \emph{\href{mailto:drozdov.pavel@spes.uniud.it}{drozdov.pavel@spes.uniud.it}, 
     \href{mailto:giorgio.gubbiotti@unimi.it}{giorgio.gubbiotti@unimi.it},   
       \href{mailto:danilo.latini@unimi.it}{danilo.latini@unimi.it}
      } }
\egroup 

\date{\today}
\numberwithin{equation}{section}

\begin{document}

\maketitle 
\begin{abstract}
    \noindent
    We determine the complete structure of the symmetry algebras associated
    with the $N$-body Calogero--Moser system and its maximally superintegrable
    discretization. We prove that the discretization naturally leads to a
    nontrivial deformation of the continuous symmetry algebra, with the
    discretization parameter playing the r\^ole of a deformation parameter.
    This phenomenon illustrates how discrete superintegrable systems can be
    viewed as natural sources of deformed polynomial algebraic structures. As a
    byproduct of these results, we also reveal a connection between the above-mentioned symmetry
    algebras and the Bell polynomials, as a consequence of the trace properties.
\end{abstract}
\tableofcontents

\section{Introduction}
\label{sec:intro}

\noindent Maximally superintegrable systems form a distinguished subclass of
finite-dimensional integrable Hamiltonian systems, characterized by the
presence of an exceptionally large number of functionally independent integrals
of motion. A classical Hamiltonian system $H(\vb q,\vb p)$ with $N$ degrees of
freedom is (Liouville) integrable if it admits $N$ functionally independent,
analytic, and single-valued constants of motion $\set{F_{1},\dots,F_{N}}$, one
of them being the Hamiltonian itself, which are mutually in involution, that
is, $\{F_i,F_j\}=0$. Under these conditions, the bounded invariant manifolds
defined by $F_i=c_i=\mathrm{const}$ with $i=1,\dots, N$ are $N$-dimensional
invariant tori, and the corresponding motions on them are conditionally
periodic~\cite{arnoldMathematicalMethodsClassical1978}. If, in addition to the
$N$ involutive integrals, there exist $k$ further functionally independent
constants of motion $\set{G_1, \dots, G_k} $ such that $1\leq k \leq N-1$, the
system is called
\emph{superintegrable}~\cite{jrClassicalQuantumSuperintegrability2013}. In this
case, the invariant tori have dimension $N-k$. The case $k=1$ defines minimally
superintegrable systems (mS), whereas the case $k=N-1$ corresponds to maximally
superintegrable systems (MS). In the MS situation, a remarkable property holds:
every bounded trajectory is closed, and the motion is strictly
periodic~\cite{nekhoroshev1972action}. If the integrals of motion are
polynomial in the canonical momenta, one speaks of \emph{polynomial
superintegrability}. Recalling that the \emph{order}, $\Ord (\hyphen)  $, of a
dynamical function is its degree in the canonical momenta, we have that a
polynomial superintegrable system is said to be $n$th-order superintegrable if
$n$ is the maximal order among a minimal generating set of constants of motion.
Classical examples of MS systems include the isotropic harmonic oscillator and
the Kepler--Coulomb model (the prototype examples of second-order MS systems),
as well as the Calogero--Moser
system~\cite{wojciechowskiSuperintegrabilityCalogeroMoserSystem1983} and the
Toda lattice~\cite{Agrotisetal2006}, among many others.
 
A distinctive feature of MS systems is that their symmetry structure is no longer encoded in an abelian algebra of first integrals. While Liouville integrability is characterized by an abelian algebra generated by the commuting integrals, MS systems possess a non-abelian symmetry algebra generated by all functionally independent
constants of motion. To obtain a closed finitely generated algebraic structure,
it is sometimes necessary to extend the generating set by including additional
linearly independent integrals. In the classical setting, the resulting
symmetry algebra is typically a \emph{polynomial Poisson algebra}, closed under
the Poisson
bracket~\cite{daskaloyannisQuadraticPoissonAlgebras2001,jrClassicalQuantumSuperintegrability2013}.
Only in exceptional cases, the algebra of integrals closes linearly, as in the
isotropic harmonic oscillator, where the symmetry algebra is $\mathfrak{su}_N$,
and in the Kepler--Coulomb system, where the symmetry algebra is
$\mathfrak{so}_{N+1}$ after a suitable rescaling of the generators for negative
energies (bounded motion). {Such polynomial algebras
    typically admit nontrivial Casimir invariants and provide a complete
    algebraic characterization of the dynamics, well beyond the abelian
    structure associated with Liouville
integrability~\cite{escobarruiz2025twodimensionalclassicalsuperintegrablesystems}.} The quadratic algebras arising in second-order superintegrable systems, and their representation theory, have been developed in depth since the early 1990s
(see~\cite{jrClassicalQuantumSuperintegrability2013, kalnins2018separation} and references
therein).

These non-abelian algebraic structures remain central in contemporary research
on superintegrability. Their study has revealed deep links between
superintegrability and the theory of special functions and orthogonal
polynomials (see~\cite{jrClassicalQuantumSuperintegrability2013} and references
therein). Moreover, they are nowadays employed to derive the spectra of quantum
systems, often avoiding direct solution of the associated Schr\"odinger
equation (see, for
example,~\cite{campoamorstursberg2025generalizedquantumzernikehamiltonians}). The quantum
associative algebras can be understood as deformations, with the deformation
parameter $\hbar$, of the underlying classical structures that are recovered in
the appropriate $\hbar \to 0$ limit.

An interesting aspect of these symmetry structures is that they can also arise
as \emph{nonlinear deformations} of an initial Lie algebra $\mathfrak{g}$. A
classical example is provided by the Higgs oscillator~\cite{Higgs1979, Leemon1979},  where in the $N=2$ case the resulting symmetry algebra is a deformation of $\su_2 $, with the deformation parameter $\kappa>0$ proportional to the scalar curvature of the underlying space.

In this paper, taking as a prototype example the Calogero--Moser
system~\cite{Calogero1971,Calogero1971erratum,MOSER1975197}, we aim to show the
existence of a novel type of deformation of polynomial symmetry algebras
associated with MS systems, obtained by considering the
discrete-time analogue of the model under investigation. In this setting, the
r\^ole of the deformation parameter is played by the lattice time spacing
$h>0$. To be more specific, the discrete-time analog of a Hamiltonian system is the theory of
maps/correspondences\footnote{A recurrence is called a map if it is single
valued and a correspondence if it is multi-valued.} of the form:
\begin{equation}
    \vb{\bar{x}} = \vb{F}(\vb{x},\vb{p}),
    \quad
    \vb{\bar{p}} = \vb{G}(\vb{x},\vb{p}),
    \quad
    \vb{\bar{x}} \coloneqq  \vb{x}(t+h),\,\,
    \vb{\bar{p}} \coloneqq  \vb{p}(t+h),
    \label{eq:qpd}
\end{equation}
for an unknown sequence $\set{(\vb{x}(t),\vb{p}(t) )}_{t\in h\Z}\subset\R^{2N}$, and preserving the canonical Poisson bracket, \ie:
\begin{equation}
    \pb*{\bar{x}_i}{\bar{x}_j} = \pb*{\bar{p}_i}{\bar{p}_j} = 0,
    \quad \pb*{\bar{x}_i}{\bar{p}_j} = \delta_{i,j}.
    \label{eq:discrpb}
\end{equation}
A  map/correspondence with this property is usually called
\emph{symplectic}, and alternatively such a
condition can be expressed by saying that the transformation
$(\vb{x},\vb{p})\longrightarrow(\vb{\bar{x}},\vb{\bar{p}})$ is canonical. This theory was initiated in the work of
Maeda~\cite{Maeda1987}.

The theory of integrability for symplectic maps, or more generally, correspondences, was developed in~\cite{Veselov1991,Bruschietal1991,Maeda1987},
see also \cite[Chap.~6]{HietarintaJoshiNijhoff2016} and the thesis
\cite[Chap.~1]{TranPhDThesis} for a comprehensive exposition of the topic.  In analogy with the continuous setting, an integral of motion for a symplectic
map/correspondence is a scalar function $I=I(\vb{x},\vb{p})$ such that it is
constant along the solutions, \ie{} $\bar{I}(\vb{x},\vb{p})\equiv
I(\vb{\bar{x}},\vb{\bar{p}})=I(\vb{x},\vb{p})$. Then, a symplectic map is
\emph{Liouville integrable} if it possesses $N$ functionally independent
invariants in involution with respect to the associated Poisson bracket. If the
map admits more functionally independent first integrals, the map
is \emph{superintegrable}, like its continuous counterpart. So, as in the continuous setting, a superintegrable symplectic map will admit the commuting integrals of motion $\Set{F_1,\ldots,F_N}$ and the non-commuting ones $\Set{G_1,\ldots,G_k}$, $1\leq k\leq N-1$.  This implies that also to discrete (maximally) superintegrable systems, one can associate a symmetry algebra, as it was discussed, for instance, in our previous
work~\cite{DGLExplicitIsomorphisms2025}.

However, there is a crucial difference between the continuous and the discrete
setting. In the continuous setting, there exists a Hamiltonian function governing the dynamics,
which is an integral of motion, while in the discrete setting, the
dynamical system is obtained by the repeated application of a canonical
transformation.  Canonical transformations do not have an associated
Hamiltonian function, but rather a generating function. In general, the
generating function is not an integral of motion of the associated discrete
dynamics. This implies that, although one can write a formal identification
between the evolution that can be derived from a generating function and
Hamilton's equations of motion, in general, an exact equivalent of the
Hamiltonian function does not exist in the discrete case. 
 This may explain why discrete integrable systems appear to be much rarer, and their structure is more rigid with respect to their continuous counterparts.

In many instances, in the limit $h\to0^{+}$ a symplectic map~\eqref{eq:qpd}
becomes a system of Hamiltonian equations. In such a case,
we say that the symplectic map~\eqref{eq:qpd} is a \emph{discretization} of the
underlying system of Hamiltonian equations. In particular, if the Hamiltonian
system is (super)integrable, and so its discretization, we will say that the
latter is a \emph{(super)integrable discretization}, see for
instance~\cite{surisProblemIntegrableDiscretization2003}.  We underline that preserving the superintegrable nature of a continuous Hamiltonian system is a very delicate task. Indeed, a symplectic map can be simply non-integrable, or integrable (but not superintegrable) discretization of a continuous superintegrable system, as highlighted in the case of the rational anisotropic caged oscillator~\cite{Evans2008} in~\cite{GubLat_sl2}. This is because, in general, the problem of
discretization is ill-posed, in the sense that it does not admit a unique solution, and that different discretizations can have different properties, see also~\cite{LeviMartinaWinternitz2015}.

In this paper, we consider the celebrated Calogero--Moser system, proven to be MS in~\cite{wojciechowskiSuperintegrabilityCalogeroMoserSystem1983}, to investigate the relationship between the symmetry algebra of a MS system and that of one of its possible MS discretizations \cite{nijhoffTimediscretizedVersionCalogeroMoser1994,ujinoAdditionalConstantsMotion2008}. 
In
particular, we seek to clarify whether the discretization can give rise to \emph{structural deformations} of a polynomial algebra. That is, if the
symmetry algebra of a MS discretization can be understood as a deformation of the original polynomial
symmetry algebra of the MS system, the latter recovered in the continuum limit $h\to0^+$. The question is motivated, for instance,
by our previous work~\cite{DGLExplicitIsomorphisms2025}, where we showed that a
class of discretizations of the harmonic oscillator, depending on a parameter,
preserves superintegrability, but only gives rise to parametric deformations in
the symmetry algebra, in the sense that the deformation can be reabsorbed into the parameters of the model. This result provided us with some interesting insights into the
discretization itself, adding, for instance, the idea of preserving not just
the integrals of motion, but also the isomorphism class of the symmetry
algebra. However, this left open the question of whether or not a maximally
superintegrable discretization could induce a structural deformation of the symmetry algebra. The results of the present paper show that the answer is affirmative: discretization can indeed give rise to structural deformations of the symmetry algebra of an MS system.

The paper is structured as follows. First, in \cref{sec:continuous-CM}, we
discuss the continuous Calogero--Moser system and its superintegrability, and we
present its symmetry algebra  $ \mathcal{A}^{(N)}$ for an arbitrary number of bodies $N$ in \cref{prop:commrelfin}, with the 
explicit low-number-of-body examples $N=2,3$ following. In \cref{sec:discrete-CM}, we proceed in a similar fashion with the Nijhoff--Pang discretization of the Calogero--Moser
model \cite{nijhoffTimediscretizedVersionCalogeroMoser1994}. We present the corresponding symmetry algebra  $\widetilde \mathcal{A}^{(N)}$ of the discrete
model in~\cref{prop:commrelfin-discr}, thereby showing \textbf{the main result of this paper: that the symmetry
algebra in the discrete case can be considered as a polynomial algebra
deformation of the continuous one}, the latter recovered when the lattice time spacing $h$ goes to zero, which is the content of~\cref{th:corol-main}.  Finally, in~\cref{sec:conclusions}, we draw
 conclusions, discuss the implications, and provide further research directions.

\section{The continuous $N$-body Calogero--Moser system and its symmetry algebra}
\label{sec:continuous-CM} 

In this section, we review the main results on the superintegrability of the
continuous Calogero--Moser model obtained in the seminal paper by S.\
Wojciechowski~\cite{wojciechowskiSuperintegrabilityCalogeroMoserSystem1983}. To
keep the discussion self-contained, we resume all the relevant properties we
will need. Then, we present the explicit form of the symmetry algebra in the
$N$-body case. The calculation of the algebra is performed in two different
steps. In the first one, we compute the Poisson commutation relations at a
formal level, since some relations fall outside the set of integrals of motion.
In the second step, using some elements of linear algebra, we show how to obtain
the missing relations and give the final form of the symmetry algebra.  We also
present as examples the symmetry algebras for the two- and the three-body cases.

\subsection{Review of the continuous model}

The Calogero--Moser system is one of the most famous integrable systems, and
one of the first to be introduced in modern times. The system describes the
dynamics of $N$ unit-mass particles moving on a line and pairwise interacting
via a potential proportional to the inverse of the squared distance. Its
history actually goes back to Jacobi, who studied the
three-body case in the XIX century~\cite{Jacobi1866}. More than a century
later, Calogero solved first the three-body quantum
case~\cite{Calogero:1969xj} and then the $N$-body
case~\cite{Calogero1971,Calogero1971erratum}. The integrability of the
classical counterpart was conjectured by Calogero in~\cite{Calogero1971} and
later proved by Moser in~\cite{MOSER1975197} using isospectral deformations. As
mentioned above, the Calogero--Moser system was shown to be maximally
superintegrable
in~\cite{wojciechowskiSuperintegrabilityCalogeroMoserSystem1983}. We note that, in modern times, although the concept of superintegrability was known at least since the $1960$s, see \eg~\cite{Fris1965,Makarov_etal1967}, the name was coined
in~\cite{wojciechowskiSuperintegrabilityCalogeroMoserSystem1983}.

The Hamiltonian function of the Calogero--Moser model is the following:
\begin{equation}\label{eq:Hamiltonian-Nbody}
	H^{(N)} = \frac{1}{2} \sum_{i=1}^N p_i^2 + {\nu^2} \sum_{1\leq i < j}^N \frac{1}{(q_i - q_j)^2 },
\end{equation}
where $\nu$ is the coupling constant parameter, which we assume to be
real.\footnote{ Note that in the literature, see \eg~\cite{surisProblemIntegrableDiscretization2003}, there exists another convention where the potential appears with the opposite sign in front of $\nu^2 $. One can switch between them using the  replacement $\nu \mapsto \imath \nu $, corresponding to repulsive and attractive potentials, respectively.   } 
For the sake of brevity, hereinafter we will
refer to the model~\eqref{eq:Hamiltonian-Nbody} as the CM model.  The Hamilton's equations of
motion read as:
\begin{align}\label{eq:Hamiltonian-ews-cont}
	\dot q_i 
	= \pb*{q_i}{ H^{(N)} }
	 =p_i, 
	\qquad 
	\dot p_i 
	= \pb*{p_i}{  H^{(N)}  } 
	= 2\nu^2 \sum_{k=1 \atop k \ne i }^N \frac{1}{(q_i - q_k )^3 }, 
	\qquad i=1,\dots,N,
\end{align}
where $\pb*{\hyphen}{\hyphen} $ is the canonical Poisson bracket,
$\pb*{q_i}{p_j} = \delta_{i,j}$.

Following the work of Moser~\cite{MOSER1975197}, we have that the Hamilton
equations~\eqref{eq:Hamiltonian-ews-cont} arise a compatibility conditions of
the following two matrices $L,M \in \gl_N (\R) $, defined by their matrix
elements as follows:
\begin{subequations} \label{eq:LaxMatrices-def-cont}
	\begin{align}
	L_{jk}&= \delta_{jk} p_j + (1- \delta_{jk} ) \frac{\imath \nu}{q_j - q_k},
	\label{eq:LaxMatixL-cont}
	\\   
	M_{jk} & = - \delta_{jk} \sum_{\ell=1 \atop \ell \ne j }^N 
	\frac{\imath \nu}{(q_j - q_\ell )^2 } 
	     + (1- \delta_{jk} ) \frac{\imath \nu}{(q_j - q_k )^2 }, 
	     \label{eq:LaxMatixM-cont}
\end{align}
\end{subequations}
under the following isospectral evolution equation:
\begin{equation}\label{eq:Lax-eq-cont}
    \dv{L}{t} = \comm{M}{L},
\end{equation}
where $\comm{\hyphen}{\hyphen}$ denotes the standard matrix commutator.
The matrices $L$ and $M$ form what is known in the literature as a \emph{Lax pair}.

The isospectrality condition~\eqref{eq:Lax-eq-cont} provides $N$ integrals of
motion from the Lax matrix $L$:
\begin{equation}\label{eq:Hs-cont}
    F_k \coloneqq \trace (L^k), \qquad k=1,\dots,N,
\end{equation}
such that $\Ord(F_{k}) = k$.   Here, we use the convention of~\cite{babelonIntroductionClassicalIntegrable2003a}. There exist other conventions, for example, where the integrals of motion are given by $ k^{-1}  \trace(L^k)$, as in~\cite{wojciechowskiSuperintegrabilityCalogeroMoserSystem1983, falquiGeometricOriginBiHamiltonian2009}. We chose the involutive integrals as in \eqref{eq:Hs-cont} because, in order to present the general formulas for the symmetry algebra, it is more convenient to have also the quantity $F_0 = N = \textrm{const}$ well-defined. 

Let us observe that the integrals of motion
$F_{k}$ are all non-local, \ie{} they depend on all dynamical variables.
Nevertheless, they are known to be functionally independent and commute between
themselves. Moreover, the Hamiltonian belongs to the above set of constants, being $F_2=2 H^{(N)}$, so that they are a \textit{bona fide} set of integrals of motion to apply Liouville--Arnold theorem.

Superintegrability of the model has been proven
in~\cite{wojciechowskiSuperintegrabilityCalogeroMoserSystem1983}, where a set
of additional integrals of motion was found. First, it is known that the
system~\eqref{eq:Hamiltonian-ews-cont} admits a second associated generalized
isospectral problem for the matrix $X$:
\begin{equation}\label{eq:eom-forX-cont}
    \dv{X}{t} = \comm{M}{X} + L \, .
\end{equation}

With the above choice of the matrix $L$~\eqref{eq:LaxMatixL-cont}, there exists
a simple solution given by $X = \diag (q_1, \dots, q_N) $. 
Note that with this choice of the matrix $X$, the symplectic structure is given by:
\begin{equation}\label{eq:sympl-structure}
	\omega = \trace(   \diff  X \wedge \diff L ) = \sum_{i=1}^N \diff q_i \wedge \diff p_i, 
\end{equation}
corresponding to the canonical Poisson bracket.
Then, it is possible to show that the quantities:
\begin{equation}
    J_{k}  \coloneqq  \trace(XL^{k-1}), \qquad k=1,\cdots,N,
    \label{eq:Jk}
\end{equation}
have a linear time evolution such that $\dv*{J_{k}}{t}= F_{k}$, and also holds $\Ord(J_{k}) = k-1$.  This allows us to construct a skew-symmetric family
of integrals of
motion~\cite{wojciechowskiSuperintegrabilityCalogeroMoserSystem1983}:
\begin{equation}\label{eq:cont-Kmk}
    K_{m,n} \coloneqq  J_m F_n -  J_n F_m, 
    \qquad 
    1\leq n<m\leq N,
\end{equation}
such that $\Ord(K_{m,n}) = m+n-1$.
Indeed, using the expressions of the $F_{k}$~\eqref{eq:Hs-cont} and
$J_{k}$~\eqref{eq:Jk}, it is easy to see that the leading order term in the momenta
never vanishes.

So, to summarize, the CM model~\eqref{eq:Hamiltonian-Nbody} admits a total of
$N(N+1)/2$ \emph{linearly independent} integrals of motion:
\begin{equation}
    \mathcal{S}^{(N)} = \set{F_{k}}_{k=1}^{N}
    \cup
    \set{K_{m,n}}_{1\leq n<m\leq N},
    \label{eq:SNdef}
\end{equation}
of which a functionally independent set is obtained by choosing the subset:
\begin{equation}
    \mathcal{S}^{(N)}_\text{f.i.} = \set{F_{k}}_{k=1}^{N}
    \cup
    \set{G_{m}}_{m=1}^{N-1},
    \qquad
    G_{m} \coloneqq 
    K_{m+1,1}.
    \label{eq:SNfidef}
\end{equation}
The following functional relations hold true:
\begin{equation}
     F_j K_{n,i} -  F_i K_{n,j} + F_n K_{i,j} = 0, \qquad i,j,n \geq 0. 
    \label{eq:funcrel}
\end{equation}
To obtain the functional relations between the elements of the set $\mathcal{S}^{(N)}$, one restricts the indices $(i,j,n)$ to the combinations with repetitions of elements in
the set $\Set{1,\ldots,N}$, up to the skew-symmetry condition. The functional
relation~\eqref{eq:funcrel}  follows trivially from rearranging the elements in the following identity:
\begin{equation}
    \det 
    \begin{pmatrix}
        F_{j} & F_{i} & F_{n}
        \\
        F_{j} & F_{i} & F_{n}
        \\
        J_{j} & J_{i} & J_{n}
    \end{pmatrix}=0.
    \label{eq:matfuncrel}
\end{equation}
In total, the functional relations~\eqref{eq:funcrel} are $C_{3}^{N}=N(N-1)(N-2)/6$, but we don't need all of them to specify the functionally dependent elements.
With the same notation as in equation~\eqref{eq:SNfidef}, we can fix the index $i=1$ and obtain:
\begin{equation}
     F_j G_{n-1} - F_1 K_{n,j}  -  F_n G_{j-1} = 0,
    \label{eq:funcrelred}
\end{equation}
which uniquely determines the integrals of motion $K_{n,j}$ with $2\leq j<n
\leq N$ in terms of  the elements of the set
$\mathcal{S}_\text{f.i}^{(N)}$.

\subsection{The symmetry algebra}
\label{subsec:cont-symmetry-algebra} 

The aim of this section is to prove that the Poisson commutation relations of
the elements of the set $\mathcal{S}^{(N)}$~\eqref{eq:SNdef} are closed.  Let
us start by showing the simplest case, \ie~the two-body case.  This is the
content of the following example:

\begin{example}[$N=2$]
    In the two-body case, the Hamiltonian function is given by:
    \begin{equation}	\label{eq:Ham-2body}
	   H^{(2)}=\frac{p_1^2+p_2^2}{2}+\frac{ \nu^2 }{(q_1-q_2)^2} \, .
    \end{equation}
    The two-body case is special among all others. As can be seen from the discussion in the above section, it is the only one where we don't have to consider linearly independent integrals of motion, but only functionally independent ones.  Indeed, following
    formulas~\eqref{eq:SNdef} and~\eqref{eq:SNfidef}, for $N=2$ we have:
    \begin{equation}
        \mathcal{S}^{(2)}  
        = 
        \mathcal{S}^{(2)}_\text{f.i} 
        = 
        \set{F_{1},F_{2},K_{2,1}},
        \label{eq:S2def}
    \end{equation}
    where we can write the integrals of motion explicitly as:
    \begin{equation}\label{eq:F1F2K21-N=2-cont-explicit} 
        F_1 = p_1+p_2, 
        \qquad 
	    F_2=2H^{(2)},
        \qquad
        K_{2,1} = ( \vb q \vdot \vb p) F_1 -  (q_1+q_2) F_2 \, .
    \end{equation}

    In this case, we can compute the Poisson commutation relations directly,
    and obtain the following quadratic algebra structure:
    \begin{equation}
      \pb{F_1}{ F_2  }=0, 
      \qquad
       \pb{F_2}{ K_{2,1}  }=0, 
     \qquad
     \pb{F_1}{K_{2,1}} = 2 F_2- F_1^2. 
        \label{eq:cont-symmetry-alg-N=2-b}
    \end{equation}
    We check that, as required by the general theory, the Hamiltonian 
     is a central element. 
    \label{ex:N2cont}
\end{example}

\begin{remark}
	We remark that the distinguishing features of the  $N=2$ model are related to the fact that it is the only separable case. Indeed, note that through the canonical lift
of the point transformation:
\begin{equation}
	Q_1 = \frac{q_1-q_2}{\sqrt{2}}, 
    \quad
    Q_2 = \frac{q_1+q_2}{\sqrt{2}}, 
\end{equation}
the Hamiltonian~\eqref{eq:Ham-2body} is mapped to:
\begin{equation}
    { H^{(2) \prime  }} =
    \frac{P_1^2+P_2^2}{2} + \frac{\nu^2}{2Q_1^2},
\end{equation}
which admits $Q_2$ as a cyclic coordinate. Moreover, this Hamiltonian
is separable with one linear and two quadratic first integrals:
\begin{equation}
	  F_1^\prime = P_2, 
    \qquad 
   F_2^\prime = {P_1^2} + \frac{\nu^2}{Q_1^2},
    \qquad
    G_1^\prime =
    Q_{2}P_{1}^2-Q_{1}P_{1}P_{2}+\nu^2 \frac{Q_{2}}{Q_{1}^2}. 
\end{equation}
These three close the quadratic algebra:
\begin{equation}
	 \pb{F_1^\prime}{F_2^\prime} = 0,
    \qquad
  \pb{F_1^\prime}{G_1^\prime} =    - F_2^\prime  ,
    \qquad
   \pb{F_2^\prime}{G_1^\prime} = F_1^\prime F_2^\prime.
\end{equation}
\end{remark}

To tackle the general $N$-body case, we will use the following auxiliary Poisson commutation 
relations of the quantities $F_{k}$ and $J_{k}$ which, in our conventions, read as: 
\begin{subequations}
    \begin{align}
        \label{eq:FFcomm}
        \pb{F_{m}}{F_{n}} &= 0, 
        \\
	\label{eq:FJ-compact} 
        \pb{J_m}{F_n} &=  n F_{m+n-2} ,
        \\
	\label{eq:known-identites-JJ}
        \pb{J_m}{J_n} &= (n-m) J_{m+n-2},
    \end{align}
    \label{eq:commFkJk}%
\end{subequations}
for $m,n>0$.
These commutation relations were obtained
in~\cite{falquiGeometricOriginBiHamiltonian2009} through the application of the
so-called necklace bracket formula for the Poisson bracket of traces. Notice that, in our conventions, the relation $\{J_1, F_1 \} =  F_0 = N $ is now included in \eqref{eq:FJ-compact}. Moreover, since $J_0 = \trace(XL^{-1}) $ is well-defined and finite, one recovers the trivial relation $\{J_1, J_1 \} =0 $ from the right-hand side of \eqref{eq:known-identites-JJ}. We will use the set of 
relations~\eqref{eq:commFkJk} to derive the Poisson commutation relations of the
$F_{k}$ and the $K_{m,n}$.

First of all, we will say that a Poisson commutation relation is \emph{formal} if on
the right hand side one can obtain either $F_{k}$ or a $K_{m,n}$ whose index
exceeds $N$, \ie~\textit{a priori} they contain elements that fall outside the set $\mathcal{S}^{(N)}$~\eqref{eq:SNdef}.  Given this definition, we have the following statement:
\begin{prop}
    The following formal Poisson commutation relations hold: 
    \begin{subequations}    \label{eq:commform}
        \begin{align}   \label{eq:FK-formal-final}
            \pb{F_i}{K_{m,n}} &=  
            i (  F_m F_{n+i-2} 
            -  F_n F_{m+i-2}), 
            \\
          \begin{split}
          \pb*{ K_{i,j}}{K_{m,n}} & = 
          F_i  ( m K_{n, j+m-2}  + n K_{j +n - 2, m}  )  
		+ F_j ( m K_{i+m-2, n}  
		  +n K_{m,i+n-2} )    
		  \\
		  &\quad  + F_m ( i K_{i+n-2, j}+ 
		  j K_{i, j+n-2} )   
		  + F_n ( i K_{j,i+m-2}
		    + j K_{j+m-2,i} )   . 
		  	    \label{eq:them:PB-KK-formal-KK}          
          \end{split}
        \end{align}
    \end{subequations}  \label{prop:commform}
\end{prop}
\begin{remark}
    We remark that putting either $i=j$ or $m=n$ in the above equations, one should obtain an identity of the form $0=0$. This is indeed true, taking into account the
    skew-symmetry of the integrals of motion $K_{m,n}$.  \label{rem:trivialcase}
\end{remark}

From the technical point of view, the proof of both sets of Poisson commutation relations follows from a judicious application of the Leibniz rule for Poisson
brackets.
\begin{proof}[\cref{prop:commform}]
	Let us prove first \cref{eq:FK-formal-final}. 
    Using the definition of $K_{m,n}$~\eqref{eq:cont-Kmk}, from the Leibniz
    rule and equation~\eqref{eq:FFcomm} we write: 
    \begin{align}
        \pb{F_i}{K_{m,n}} = 
         \pb{ F_i}{J_m F_n} - \pb{F_i}{J_n F_m}
		=   \{ F_i, J_m \} F_n -   \{F_i, J_n \} F_m.   
    \end{align}
    Using then the equation~\eqref{eq:FJ-compact}, one arrives at the expression
    in~\eqref{eq:FK-formal-final}. 
    We remark that the Poisson commutation relations $\pb{F_{i}}{K_{m,1}}$ was already
    obtained with a different method
    in~\cite{wojciechowskiSuperintegrabilityCalogeroMoserSystem1983}.
    \label{rem:wojciechowskicomm}

Let us proceed to \cref{eq:them:PB-KK-formal-KK}. 
    Define $ R_{m,n} \coloneqq  J_m F_n  $, such that:
    \begin{equation}\label{eq:thmKKformal-K-in-terms-of-Rs}
        K_{m,n} = R_{m,n} - R_{n,m} \eqqcolon R_{[m,n]}, 
    \end{equation}
    where the square brackets stand for skew-symmetrization. Note that by construction: 
  	\begin{equation}\label{eq:Rab-prop}
            R_{i,j} F_k = F_j  R_{i,k} , 
            \qquad 
            i,j,k>0. 
    \end{equation}

    Following \cref{rem:trivialcase}, we can assume, without loss of
    generality, that $i\neq j$ and $m\neq n$. Introduce a tensor $ {\mR_{i,j,k,l}
    \coloneqq \{ R_{i,j} , R_{k,l} \}} $. We then have
    by~\eqref{eq:thmKKformal-K-in-terms-of-Rs}: 
    \begin{equation}
    \{ K_{i,j}, K_{m,n} \} = 
              \mR_{i,j,m,n} -  \mR_{i,j,n,m} -  \mR_{j,i,m,n} +  \mR_{j,i,n,m}
              =  \mR_{i,j \, [m,n]} - \mR_{ j,i \, [m,n] } 
              \eqqcolon \mR_{ [i,j] \, [m,n] }. 
    \end{equation} 
    Let us now proceed to the calculation of ${\mR_{i,j,k,l}}$. Using the definition
    of $ R_{i,j} $ and the Leibniz rule, we obtain: 	
   \begin{align} 
       	\mR_{i , j , k,  l}  
            &=   \{ J_i F_j , J_k F_l \} 
            =     
             - F_l J_i  \{ J_k, F_j \} 
             + F_j F_l \{ J_i, J_k \} 
             + F_j J_k \{ J_i, F_dl \} 
              \notag 
            \\
              	 & =  
            (k-i) F_j F_l J_{i+k-2} 
            + l F_j    F_{i+l-2}  J_k   
            - j F_l   F_{j+k-2} J_i 
             	\notag
              \\ 
              & = - j F_l R_{i, j+k -2}
              + F_j \bigl( 
              (j-i) R_{i+k-2, l} + l R_{k, i+l-2}
              \bigr). 
            \label{eq:mR-expanded} 
    \end{align}
       Performing an antisymmetrization with respect to $i,j $ and $k,l$,  and doing some algebra, one arrives at
    the expression~\eqref{eq:them:PB-KK-formal-KK}.    
\end{proof}
Now, a crucial point is that the indices in the Poisson commutation relations~\eqref{eq:commform} can exceed
$N$. Nevertheless, the symmetry algebra is closed since all the extra terms can
be expressed via the elements of the set $\mathcal{S}^{(N)}$~\eqref{eq:SNdef}. Note that the elements $K_{j,i} $ with $j<i$ can always be rewritten using the antisymmetry condition as $K_{j,i} = -K_{i,j} $. Then,  at a fixed $N$, and analyzing the right-hand side of the Poisson commutation 
relations~\eqref{eq:commform}, one finds that we need to determine the following additional functionally dependent first integrals:
\begin{subequations}\label{eq:FandKneed-cont}
    \begin{gather}
        \label{eq:FNsneed}
        \begin{array}{cccc}
             F_{N+1},& F_{N+2},& \ldots, & F_{2N-2},
        \end{array}
        \\
        \begin{array}{ccccc}
              K_{1,0}, &   K_{ 2,0 }, & \ldots, 
              &  K_{N-1, 0}, &   K_{N, 0}, 
             \end{array}  
             \label{eq:K0-row}
             \\ 
        \label{eq:KNspneed}
        \begin{array}{ccccc}
             K_{N+1,1}, & K_{N+1,2}, & \ldots, & K_{N+1,N-1}, & K_{N+1,N},
             \\
             K_{N+2,1}, & K_{N+2,2}, & \ldots, & K_{N+2,N-1}, & K_{N+2,N},
             \\
             \vdots & & & & \vdots 
             \\
             K_{2N-3,1}, & K_{2N-3,2} ,& \ldots, & K_{2N-3,N-1}, & K_{2N-3,N},
             \\
             K_{2N-2,1}, & K_{2N-2,2}, & \ldots, & K_{2N-2,N-1}.
        \end{array}
    \end{gather}
\end{subequations}
Abstractly, the functions $F_k$, $J_k$, and $K_{m,n}$ can be defined for
arbitrary values of the indices  $k,m,n \geq  0$, so the above expressions make perfect sense. 
First, let us discuss the set $K_{1,0}, \ldots, K_{N,0} $. Analyzing the right-hand side of \eqref{eq:them:PB-KK-formal-KK}, given $i,j,m,n >0 $ for the functions belonging to $\mathcal{S}^{(N)} $,  we note that the terms \eqref{eq:K0-row} can appear if at least one of the following equalities holds:
\begin{subequations}\label{eq:equalities-ijmn2}
	\begin{align}
		j+ m - 2 & =0 \quad \Longrightarrow \quad j=1, \ m=1,
		\\
		j+ n -2 & = 0 \quad \Longrightarrow \quad j=1, \ n=1, 
		\\
		i+ m - 2 & =0 \quad \Longrightarrow \quad i=1, \ m=1,
		\\
		i+ n - 2 &= 0 \quad \Longrightarrow \quad i=1, \ n=1. 
	\end{align}
\end{subequations}  
Let us observe that, fixing an index $n=0 $ in functional relation  \eqref{eq:funcrel}, we can write:  
\begin{equation}\label{eq:funcrel-K0}
	    F_j K_{0,i} -  F_i K_{0,j} = N K_{ji}.  
\end{equation} 
In this way, one is able to express all the products containing $K_{i,0}$. 

To proceed with other rows in \eqref{eq:FandKneed-cont}, let us note that  both $F_k$ and $J_k$ are related to the trace of
powers of the Lax matrix. So, following an idea suggested already
in~\cite{falquiGeometricOriginBiHamiltonian2009}, if $k>N$, we use the
Cayley--Hamilton theorem, see \eg~\cite{loehrAdvancedLinearAlgebra2024}, to
express them in terms of elements of $\mathcal{S}^{(N)}$~\eqref{eq:SNdef}.

Following the results reported in \cref{app:symmpoly} (see
formula~\eqref{eq:app-arbitrary-matr-traces-Bell}), we can express $F_{N+s} $ through the Bell polynomials evaluated on the power sums, and hence in terms of the traces as:
\begin{equation}
	 F_{N+s} =  -  \sum_{k=1}^N 
    \frac{1}{k!} B_k 
	\bigl( - 0! F_1, -1! F_2, -2! F_3, \ldots, 
	-(k-1)! F_k  \bigr) F_{N+s-k}.
        \label{eq:FNs}
\end{equation}
Note that this is not enough yet, since this last formula expresses
$F_{N+s}$ in terms of $F_{1},\ldots,F_{N+s-1}$. However,
it is easy to see that the $F_{N+s}$ can be determined recursively starting
from $F_{N+1}$. Since, as highlighted in formula~\eqref{eq:FNsneed}, we need 
$\set{F_{N+s}}_{s=1}^{N-2} $, it means that to express everything in terms of $F_{1}, \ldots, F_{N}$, one solves the following triangular system:
\begin{equation}
    \left\{
        \begin{aligned}
	 F_{N+1} &=  -  \sum_{k=1}^N 
    \frac{1}{k!} B_k 
	\bigl( - 0! F_1, -1! F_2, -2! F_3, \ldots, 
        -(k-1)! F_k  \bigr) F_{N+1-k},
            \\
	 F_{N+2} &=  -  \sum_{k=1}^N 
    \frac{1}{k!} B_k 
	\bigl( - 0! F_1, -1! F_2, -2! F_3, \ldots, 
        -(k-1)! F_k  \bigr) F_{N+2-k},
        \\
        &\vdots
        \\
	 F_{2N-2} &=  - \sum_{k=1}^N 
    \frac{1}{k!} B_k 
	\bigl( - 0! F_1, -1! F_2, -2! F_3, \ldots, 
	-(k-1)! F_k  \bigr) F_{2N-2-k}.
        \end{aligned}
    \right.
        \label{eq:FNstriang}
\end{equation}
This expression solves the problem completely. Moreover, we observe that,
by induction and since $\deg_{\vb{y}} B_k(\vb{y})=k$, it follows that: 
\begin{equation}
    \deg_{\mathcal{S}^{(N)}}F_{N+s} = N+s,
    \quad
    s=1,\ldots,N-2. 
    \label{eq:degreeFNp}
\end{equation}
We also notice that one can compare formula~\eqref{eq:FNstriang} with an 
analogous one presented for the Toda lattice in~\cite[\S 3]{Agrotisetal2006}.

In the same way, we can multiply on the left the
relation~\eqref{eq:matrix-pol-CHthm} reported in \cref{app:symmpoly}, for $A=L $, by the matrix $XL^{s-1}$ for $s\in\N$. Given that $\trace (XL^{k-1}) = J_k$, 
taking the traces, we obtain:
\begin{equation}\label{eq:JNstraces}
    J_{N+s} =
    \sum_{i=1}^N {(-1)}^{i+1} e_i (\vb* \lambda) J_{N+s-i}.
\end{equation}
Hence, fixing $s\in\N$, we can write:
    \begin{equation}
        K_{N+s, \ell}
        =
        -\sum_{k=1}^{N}\frac{1}{k!} 
        B_k \bigl( - 0! F_1, -1! F_2, -2! F_3, \dots, -(k-1)! F_k \bigr)
        K_{N+s-k,\ell}.
         \label{eq:KNsNp}
    \end{equation}

As before, it is not enough yet, since this last formula expresses
$K_{N+s,\ell }$ in terms of the elements of
$\mathcal{S}^{(N)}$~\eqref{eq:SNdef}.  Yet again, it is possible to see that
for a fixed $\ell $, the function $K_{N+s,\ell}$ can be determined recursively
starting from $K_{N+1,\ell}$. Since, as highlighted in
formula~\eqref{eq:KNspneed}, we need the functions $\set{K_{N+s,\ell}}_{s=1}^{N-2}$, for $\ell \in
\Set{1,\ldots,N-1}$, and the functions $\set{K_{N+s,N}}_{s=1}^{N-3}$.  That is, we have to solve the following families of triangular systems:
\begin{subequations}\label{eq:KNsltottriang} 
\begin{equation}
    \left\{
        \begin{aligned}
	       K_{N+1,\ell }
        &=-
        \sum_{k=1}^{N}\frac{1}{k!} 
        B_k \bigl( - 0! F_1, -1! F_2, -2! F_3, \dots, -(k-1)! F_k \bigr)
        K_{N+1-k,\ell}
            \\
	       K_{N+2,\ell}
        &=-
        \sum_{k=1}^{N}\frac{1}{k!} 
        B_k \bigl( - 0! F_1, -1! F_2, -2! F_3, \dots, -(k-1)! F_k \bigr)
        K_{N+2-k,\ell },
        \\
        &\vdots
        \\
	    K_{2N-2,\ell}
        &=-
        \sum_{k=1}^{N}\frac{1}{k!} 
        B_k \bigl( - 0! F_1, -1! F_2, -2! F_3, \dots, -(k-1)! F_k \bigr)
        K_{2N-2-k,\ell},
        \end{aligned}
    \right.
        \label{eq:KNsltriang}
\end{equation}
for $\ell \in \Set{1,\ldots,N-1}$, and
\begin{equation}
    \left\{
        \begin{aligned}
	       K_{N+1,N}
        &=-
        \sum_{k=1}^{N}\frac{1}{k!} 
        B_k \bigl( - 0! F_1, -1! F_2, -2! F_3, \dots, -(k-1)! F_k \bigr)
        K_{N+1-k,N}
            \\
	       K_{N+2,N}
        &=-
        \sum_{k=1}^{N}\frac{1}{k!} 
        B_k \bigl( - 0! F_1, -1! F_2, -2! F_3, \dots, -(k-1)! F_k \bigr)
        K_{N+2-k,N},
        \\
        &\vdots
        \\
	    K_{2N-3,N}
        &=-
        \sum_{k=1}^{N}\frac{1}{k!} 
        B_k \bigl( - 0! F_1, -1! F_2, -2! F_3, \dots, -(k-1)! F_k \bigr)
        K_{2N-2-k,N}.
        \end{aligned}
    \right.
        \label{eq:KNsNtriang}
\end{equation}    
\end{subequations}
These last two expressions solve the problem completely. Moreover, we observe 
that by induction, taking into account that $\deg_{\vb{y}} B_k(\vb{y})=k$, it follows that:
\begin{equation}
    \begin{array}{lll}
        \deg_{\mathcal{S}^{(N)}}K_{N+s,1} = N+s-1,
        & s=1,\ldots,N-2,
        \\
        \deg_{\mathcal{S}^{(N)}}K_{N+s,\ell} = N+s,
        & s=1,\ldots,N-2,
        & \ell =2,\ldots,N-1,
        \\
        \deg_{\mathcal{S}^{(N)}}K_{N+s,N} = N+s,
        & s=1,\ldots,N-3, 
    \end{array}
    \label{eq:degreeKNpltot}
\end{equation}
where we used that for $\ell=1$, since $K_{1,1}=0$, the highest degree 
Bell polynomial is annihilated.

Therefore, the set of these rules allows us to write explicitly the form of the
symmetry algebra of the CM model~\eqref{eq:Hamiltonian-Nbody}. We enclose this
result in the following proposition:

\begin{prop}
    The symmetry algebra $\mathcal{A}^{(N)}$  of the $N$-body CM model is generated by
    the elements of the set $\mathcal{S}^{(N)}$~\eqref{eq:SNdef} with 
    Poisson commutation relations given in \cref{eq:commform},
    provided the additional closure relations given in
    \cref{eq:FNstriang,eq:KNsltottriang}.  Moreover, $\mathcal{I}^{(N)} \coloneqq  \langle
    F_i\rangle_{i=1}^{N}$ is an ideal of ~$\mathcal{A}^{(N)}$. Finally, for
    $N>2$ the degree of the polynomial symmetry algebra $\mathcal{A}^{(N)}$ is $2N-1$, while  $\mathcal{A}^{(2)}$ is quadratic.\label{prop:commrelfin}
\end{prop}

\begin{proof}
Everything in the above proposition was proved, except for the last statement
about the degree of $\mathcal{A}^{(N)}$ for $N>2$ (the case $N=2$ was discussed
in \cref{ex:N2cont}). We have:
\begin{equation}
    \deg\mathcal{A}^{(N)}=
    \max\Set{\max_{\substack{i=1,\ldots,N\\1\leq n<m\leq N}}\deg_{\mathcal{S}^{(N)}}\pb{F_i}{K_{m,n}},
    \max_{\substack{1\leq j<i\leq N\\1\leq n<m\leq N}}\deg_{\mathcal{S}^{(N)}}\pb*{K_{i,j}}{K_{m,n}}}.
\end{equation}
We observe that from~\cref{eq:FK-formal-final}, taking into account~\cref{eq:degreeFNp,eq:degreeKNpltot}, we have:
\begin{equation}\label{eq:PB-deg-estimate}  
    \max_{\substack{i=1,\ldots,N\\1\leq n<m\leq N}}\deg_{\mathcal{S}^{(N)}}\pb{F_i}{K_{m,n}}
    =
    \deg_{\mathcal{S}^{(N)}}\pb{F_{N}}{K_{N,N-1}}
    =
    \deg_{\mathcal{S}^{(N)}}F_{N-1}F_{2N-2} = 2N-1.
\end{equation}
In the same way, from~\cref{eq:them:PB-KK-formal-KK} and
taking into account~\cref{eq:degreeFNp,eq:degreeKNpltot}, we have:
\begin{equation}
\begin{split}
	 \max_{\substack{1\leq j<i\leq N\\1\leq n<m\leq N}}\deg_{\mathcal{S}^{(N)}}\pb*{K_{i,j}}{K_{m,n}}
    &=
    \deg_{\mathcal{S}^{(N)}}\pb{K_{N,N-2}}{K_{N,N-1}}
    \\
    &=
    \deg_{\mathcal{S}^{(N)}}F_{N-2}K_{2N-2,N-2} = 2N-1.
\end{split}	
\end{equation}
Hence, in the end we obtain $\deg\mathcal{A}^{(N)}=2N-1$, finishing the proof
of the proposition.
\end{proof}

We conclude this subsection with a detailed study of the three-body case. 

\begin{example}[$N=3$]\label{example:cont-N=3}
    Let us consider the CM model in the three-body case, \ie{} the Hamiltonian
    system in~\cref{eq:Hamiltonian-Nbody} with $N=3$:
    \begin{equation}
	   H^{(3)}=\frac{p_1^2+p_2^2+p_3^2}{2}
	   +\nu^2 \left[ \frac{1}{(q_1-q_2)^2}+\frac{1}{(q_1-q_3)^2}+\frac{1}{(q_2-q_3)^2} \right].
	   \label{eq:Ham1}
    \end{equation}
    Following formulas~\eqref{eq:SNdef} and~\eqref{eq:SNfidef}, for $N=3$, the system is endowed with the following linearly independent integrals of motion:
    \begin{equation}
        \mathcal{S}^{(3)} = \Set{F_1,F_2,F_3,K_{2,1},K_{3,1},K_{3,2}}.
    \end{equation}
    From~\eqref{eq:Hs-cont} and~\eqref{eq:cont-Kmk}, these integrals of motion have the following explicit expressions:
    \begin{subequations}\label{eq:constants3DcaseF}
    \begin{align}
	   F_1 &=p_1+p_2+p_3, 
	   \\ 
	   F_2 & = 2 H^{(3)}, 
	   \\ 
	   F_3 & =  \left(p_1^3+p_2^3+p_3^3  \right) 
	   +3 \nu^2 \left[ \frac{p_1+p_2}{(q_1-q_2)^2}+\frac{p_1+p_3}{(q_1-q_3)^2}+\frac{p_2+p_3}{(q_2-q_3)^2} \right],
	\end{align}
	\end{subequations}
	and:
	  \begin{subequations}\label{eq:constants3DcaseK}
	  \begin{align}
       K_{2,1} & =  (\vb q \vdot \vb p)  F_1 -   (q_1+q_2+q_3) F_2, 
	   \\
	   K_{3,1} & = (\vb q \vdot \vb A ) F_1 -   (q_1+q_2+q_3) F_3, 
        \\
        K_{3,2} &=   ( \vb q \vdot \vb A)  F_2 -  ( \vb q \vdot \vb p ) F_3,
    \end{align}	
    \end{subequations}
    where the vector $\vb A \coloneqq (A_1, A_2, A_3) $ is defined through its
    components as: 
    \begin{equation}\label{eq:vecA-def}
	A_i \coloneqq p_i^2 + { \sum_{j=1 \atop j \ne i }^3} \frac{\nu^2}{(q_i - q_j )^2},
	\qquad 
        {i=1,2,3}.
    \end{equation}
    Note that $A_1 +A_2 +A_3 = 2 H^{(3)} $.

    We can write down the Poisson commutation relations of the algebra $\mathcal{A}^{(3)}$ 
    generated by~$\mathcal{S}^{(3)}$. First, there is an abelian part:
    \begin{equation}
        \pb{ F_2}{\hyphen }=0,
        \qquad
        \pb{F_1}{F_3}=0.
        \label{hamcentral}
    \end{equation}
    Then, there are the following formal Poisson commutation relations from~\cref{eq:FK-formal-final}:
    \begin{equation}
        \begin{array}{lll}
           \pb{F_1}{K_{2,1}}  = -F_{1}^2+3F_{2}, 
	       &
	       \pb{F_1}{K_{3,1}}  = 3 F_{3}-F_{1}F_{2},
	       &
	       \pb{F_1}{K_{3,2}}  =  F_{1} F_{3}- F_{2}^2,
           \\
	       \pb{F_3}{K_{2,1}}  = 3(F_{2}^2-F_{1}F_{3}), 
	       &
	       \pb{F_3}{K_{3,1}}  = 3(F_{2} F_{3}- F_{1}F_{4}),
	       &
	       \pb{F_3}{K_{3,2}}  = 3(F_{3}^2-F_{2}F_{4}),
        \end{array}
           \label{eq:cont-FK3}
    \end{equation}
    and the ones from~\eqref{eq:them:PB-KK-formal-KK}:
    \begin{equation}\label{eq:cont-KK3}
        \begin{aligned}
	       \pb{K_{2,1}}{K_{3,1}} & = 
	       2 F_1 K_{3,1}  
	       -F_2 ( 3K_{2,1} + K_{3,0})
	       + F_3 K_{2,0}, 
	       \\
	       \pb{K_{2,1}}{K_{3,2}} & = 5F_{1}K_{3, 2} 
	       - 4 F_2  K_{3,1} + 3 F_3 K_{2,1}, 
	       \\
	       \pb{K_{3,1}}{K_{3,2}} & = 
	        3 F_1 K_{4,2} - F_2 ( K_{3,2} + 3 K_{4,1} ) 
	       +  2 F_3 K_{3,1}. 
      \end{aligned}    
    \end{equation}
    Here, by \cref{eq:funcrel-K0}, we have $ F_3 K_{2,0} - F_2 K_{3,0} =-  3 K_{3,2}  $. Furthermore, as observed above, to determine the actual Poisson commutation
    relations, we need to provide the form of $F_4$, $K_{4,1}$, and $K_{4,2}$.
    This is readily given using equations~\eqref{eq:FNs} and~\eqref{eq:KNsNp}:
    \begin{subequations}
        \begin{align}
            F_4 &= \frac{4}{3} F_{1}F_{3} - F_{1}^2F_{2} +\frac{1}{2}F_{2}^2 
            +\frac{1}{6}F_{1}^4,
            \\
            K_{4,1} &= F_{1}K_{3, 1}
            - \frac{1}{2} \left( F_{1}^2- F_{2}\right) K_{2, 1},
            \\
            K_{4,2} &= F_{1}K_{3, 2} 
            - \frac{1}{6} \left( F_{1}^3 -3 F_{1}F_{2}+2F_{3}\right)K_{2, 1}.
        \end{align}
        \label{eq:F4K41K42}
    \end{subequations}
    Observe that, as expected, $\deg_{\mathcal{S}^{(3)}}F_4 = 
    \deg_{\mathcal{S}^{(3)}}K_{4,2}=4$ and $\deg_{\mathcal{S}^{(3)}}K_{4,1}=3$.
    Then, by substituting equation~\eqref{eq:F4K41K42} into formulas~\eqref{eq:cont-FK3} 
    and~\eqref{eq:cont-KK3}, we obtain the explicit form of the symmetry algebra for the three-body CM system, \ie{}:
    \begin{subequations}
        \begin{align}
            \pb{F_1}{K_{2,1}}  &= -F_{1}^2+3F_{2}, 
            \label{eq:F1K21N3}
            \\
	           \pb{F_1}{K_{3,1}}  &= -F_{1}F_{2}+3F_{3},
               \label{eq:F1K31N3}
               \\
	           \pb{F_1}{K_{3,2}}  &=  F_{1} F_{3}-  F_{2}^2,
               \label{eq:F1K32N3}
               \\
	           \pb{F_3}{K_{2,1}}  &= 3(F_{2}^2 -F_{1}F_{3} ), 
               \label{eq:F3K21N3}
	           \\
	           \pb{F_3}{K_{3,1}}  &= 
               3F_{2} F_{3}-4F_{1}^2F_{3} +3F_{1}^3 F_{2} - \frac{3}{2} F_{1}F_{2}^2 
            - \frac{1}{2}F_{1}^5,
            \label{eq:F3K31N3}
	           \\
	          \pb{F_3}{K_{3,2}}  &= 
               3F_{3}^2-4F_{1}F_{2}F_{3} +3 F_{1}^2 F_{2}^{2} 
               -\frac{3}{2} F_{2}^3 
            -\frac{1}{2}F_{1}^4F_{2},
            \label{eq:F3K32N3}
            \\
            \pb{K_{2,1}}{K_{3,1}}  &= 2F_{1}K_{3, 1}-3F_{2}K_{2, 1}-3K_{3, 2},
            \label{eq:K21K31N3}
                \\
                \pb{K_{2,1}}{K_{3,2}}  &= 5F_{1}K_{3, 2} -4F_{2}K_{3,1} 
           +3F_{3}K_{2, 1},
           \label{eq:K21K32N3}
	       \\
	       \begin{split}
	       	 \pb{K_{3,1}}{K_{3,2}}  &= 2F_3K_{3, 1}
	       	 +3F_1^2K_{3, 2} 
	       	 -\frac{1}{2}F_{1}^4K_{2, 1}
           +3 F_{1}^2F_{2} K_{2,1}
           \\
           & \quad  - F_{1}F_{3}K_{2, 1}
           -3F_{1}F_{2}K_{3, 1}
           -\frac{3}{2} F_{2}^2K_{2, 1}-F_{2}K_{3, 2},
              \label{eq:K31K32N3} 
	       \end{split}    
        \end{align}
    \end{subequations}
 together with \eqref{hamcentral}. Observe that, as expected, the algebra $\mathcal{A}^{(3)}$ has degree five:
    the Poisson commutation relations~\eqref{eq:F3K31N3}, \eqref{eq:F3K32N3},
    and~\eqref{eq:K31K32N3} have all degree five in $\mathcal{S}^{(3)}$. Recall
    that $K_{3,2}$ is not functionally independent (but linearly independent)
    on other integrals of motion. Following~\eqref{eq:funcrelred}, the
    functional relation linking $K_{3,2}$ with the functionally independent
    invariants are given by: 
    \begin{equation}
    	 F_2 K_{3,1} - F_1 K_{3,2} 
    	- F_3 K_{2,1} =0. 
    \end{equation}  
\end{example}

\section{The Nijhoff--Pang discretization of the $N$-body Calogero--Moser system and its symmetry algebra}
\label{sec:discrete-CM}

In this section, we first review the main results
of~\cite{nijhoffTimediscretizedVersionCalogeroMoser1994,
ujinoAdditionalConstantsMotion2008} concerning the dynamics and integrability
of a MS discretization of the CM model. Then, we proceed
by computing the corresponding symmetry algebra for the arbitrary $N$-body case in the same fashion
as for the continuous case considered in~\cref{sec:continuous-CM}. We thereby
show the main result of the paper: how discretization leads to a deformation
of the polynomial symmetry algebra of the model with respect to the
discretization parameter~$h$. We also present, as explicit examples, the deformed symmetry
algebra for the two- and the three-body cases.

\subsection{Review of the discrete model}

In~\cite{nijhoffTimediscretizedVersionCalogeroMoser1994}, Nijhoff and Pang
introduced an integrable discretization of the CM system. Later, this discretization was proved to be MS
in~\cite{ujinoAdditionalConstantsMotion2008},  building a discrete analog of the
$G_j$~(see \cref{eq:SNfidef}). Hereinafter, we will address this discrete
system as Nijhoff--Pang discretization of the $N$-body CM system, or in short
NPdCM system. The dynamics of the NPdCM system is governed by the following
difference equations of motion in canonical form (see~\cref{eq:qpd}):
\begin{subequations}\label{eqs:discrete-eoms-canonical} \begin{align} p_k & =
        \frac{c_0}{h} \left[ 1 - c_0 \sum_{j=1}^N \frac{1}{ \bar x_j - x_k +c_0
            } + c_0 \sum_{j=1 \atop j \ne k }^N \frac{1}{x_j - x_k} \right], \\
            \bar p_k & =  \frac{c_0}{h} \left[ 1 - c_0 \sum_{j=1}^N
            \frac{1}{\bar x_k - x_j +c_0 } +  c_0 \sum_{j=1 \atop j \ne k }^N
        \frac{1}{\bar x_k - \bar x_j} \right],  \end{align}
    \end{subequations} where $h>0 $ is a discretization parameter, and $c_0 $
is defined such that $c_0^2 \coloneqq  - \imath h \nu $.  Through the rescaling
$x_{k}(t)=q_{k}(t)+h t$, the system~\eqref{eqs:discrete-eoms-canonical}
constitutes a discrete version of the continuous equations of
motion~\eqref{eq:Hamiltonian-ews-cont}. 

Similarly to the continuous case, one may find a Lax pair. This can be done by
performing a discretization of the isospectral evolution equation
\eqref{eq:Lax-eq-cont}, which can be written as: 
\begin{equation}\label{eq:Lax-eq-discr}
    \bar L \widetilde M = \widetilde M L, 
\end{equation}
where the matrices $L, \widetilde M \in \gl_N (\R) $ are defined by their
matrix elements as follows: 
\begin{subequations} \label{eq:LaxMatrices-def-discr}
	\begin{align}
	L_{jk}&= \delta_{jk} p_j + (1- \delta_{jk} ) \frac{\imath \nu}{x_j - x_k},
	\label{eq:LaxMatixL-discr}
	\\   
	 \M_{jk} & = \frac{c_0}{\bar x_j - x_k + c_0 }. 
	     \label{eq:LaxMatixM-discr}  
\end{align}
\end{subequations}
We remark that the matrix $L$ can be chosen to be of the same form as in the continuous case, \cf~\eqref{eq:LaxMatrices-def-cont}. Therefore, it provides us with the $N$ integrals of motion, given by the same expression as in the continuous case~\eqref{eq:Hs-cont}.

In turn, a generalization of the Wojciechowski construction~\eqref{eq:cont-Kmk}
is nontrivial. First, let us notice that the discrete version of the
generalized isospectral problem \eqref{eq:eom-forX-cont} is provided by: 
\begin{equation}\label{eq:eom-forX-discr}
	\bar{X} \M = \M X + h  \M L ( \1 - h c_0^{-1} L)^{-1},
\end{equation}
and the simplest solution is again  $X = \diag \left( x_1, \dots, x_N \right) $.
Therefore, the modified integrals of motion for the discrete case,
corresponding to~\eqref{eq:cont-Kmk}, can be defined as follows:
 \begin{equation}\label{eq:discr-Kmk}
	\K_{m,n} \coloneqq  \trace \left[ X \left( \1 - h c_0^{-1} L \right) L^{m-1} \right] \trace \left( L^n \right) 
	   - \trace \left( L^m \right) \trace \left[ X \left( \1 - h c_0^{-1} L \right)  L^{n-1} \right], 
	  \quad 
    1\leq n<m\leq N.
\end{equation}
One can observe that in the continuum limit $h \to 0 $, we recover the integral of motion~$K_{m,n} $~\eqref{eq:cont-Kmk} from the continuous case. Note that here we have  $\Ord ( \K_{m,n} ) = m + n$.
Indeed, from \cref{eq:Lax-eq-discr} and
\eqref{eq:eom-forX-discr}, we have:
\begin{equation}
	\bar X (  \1 - h c_0^{-1} \bar L ) = 
	\M \bigl[ X  (\1 - h c_0^{-1} L ) + h L  \bigr] \M^{-1}. 
\end{equation} 
We use the cyclicity of trace together with a property of similarity
transformation:
\begin{equation}
    ( \overline L )^n  = ( \M L \M^{-1} )^n = \M L^n \M^{-1} 
    \implies
    \tr [  \overline{L}^n  ] 
    = \tr [ L^n ].  
\end{equation}
Thus, using linearity and cyclicity of trace again, one can prove that $\K_{m,n}$ is conserved:\footnote{We remark that the proof for the subset of the functionally independent integrals of motion, \ie{} for $ \set{ G_{m} =  K_{m+1,1}}_{m=1}^{N-1}  $ was done in~\cite{ujinoAdditionalConstantsMotion2008}.}  
\begin{align}
	\bar \K_{m,n} & = \trace \left[ \bar X \left( \1 - h c_0^{-1} \bar L 
	  \right) \bar L^{m-1} \right] \trace \left( \bar L^n \right) 
	   - \trace \left( \bar L^m \right) \trace \left[ \bar X \left( \1 - h c_0^{-1} \bar L \right) \bar L^{n-1} \right]
	   \notag
	   \\
	   & = \trace \left[ 
	   X  L^{m-1} - h c_0^{-1} X L^{m}    + h   L^{m}      \right] \trace (  L^n) 
	   - \trace \left(  L^m \right) \trace \left[    X L^{n-1} 
	   - h c_0^{-1} X  L^{n}   
	   + h  L^{n}    \right]
	   \notag
	   \\
	    & = \trace \left[ 
	   X  ( \1 - h c_0^{-1}  L)  L^{m-1}    
	      \right] \trace (  L^n) 
	   - \trace \left(  L^m \right) 
	   \trace \left[  
	   X ( \1 
	   - h c_0^{-1}   L )  L^{n-1} 
	     \right]
	     \notag
	     \\
	     & = \K_{m,n}. 
\end{align}
Therefore, the NPdCM model admits a total of $N(N+1)/2$ \emph{linearly independent} integrals of motion:
\begin{equation}\label{eq:SNdef-discr}
     \widetilde \mathcal{S}^{(N)} = \set{F_{k}}_{k=1}^{N}
    \cup
    \set{\K_{m,n}}_{1\leq n<m\leq N},
\end{equation}
of which a functionally independent subset is given by:
\begin{equation}\label{eq:SNfidef-discr}
   \widetilde  \mathcal{S}^{(N)}_\text{f.i.} = \set{F_{k}}_{k=1}^{N}
    \cup
    \set{\G_{m}}_{m=1}^{N-1},
    \qquad
    \G_{m} \coloneqq 
    \K_{m+1,1}, 
\end{equation}
\cf~\cref{eq:SNdef}  and \cref{eq:SNfidef}. 

By construction,  upon replacing $K_{i,j}
\mapsto \K_{i,j}$,
the same functional relations as in the continuous case (see \cref{eq:funcrel})
hold true:
\begin{equation}
     F_j \K_{n,i} -  F_i \K_{n,j} + F_n \K_{i,j} = 0, \qquad i,j,n \geq 0,  
    \label{eq:funcrel-discr}
\end{equation}
where the same remark we made in the continuous setting also applies here. 

Finally, let us observe that the
modified integrals of motion~$\K_{m,n}$ can be rewritten as follows: 
\begin{equation}\label{eq:Kdiscr-via-flow}
    \K_{m,n} = K_{m,n} - h c_0^{-1} K^{(1)}_{m+1, n+1}
    = K_{m,n} + \alpha \sqrt{h} K^{(1)}_{m+1, n+1},
    \qquad
    \alpha \coloneqq  
    \frac{1+\imath }{\sqrt{2 \nu }} ,
\end{equation}
where $K_{m,n} $ is an integral of motion in the continuous
case~\eqref{eq:cont-Kmk}, and $K^{(1)}_{m+1, n+1} $ is the first member of the so-called
higher flow: 
\begin{equation}\label{eq:genflowa1} 
	K^{(a)}_{m, n}=  J_m F_{a+n-2} - J_n F_{a+m-2}, 
\end{equation}
reported in \cite{wojciechowskiSuperintegrabilityCalogeroMoserSystem1983}.  
Notice that $ \Ord (K^{(a)}_{m,n} ) = a+n+m-3$
for $ m \neq n$, and $K^{(2)}_{m,n} \equiv K_{m,n}$. 

\subsection{The symmetry algebra}

The aim of this section is to obtain the expressions defining the symmetry
algebra of the NPdCM system. We do it in the same fashion as for the continuous
case (see~\cref{subsec:cont-symmetry-algebra}). We also present the main result
of the paper, namely the derivation of the symmetry algebra of the discretized model as a deformation of its continuous counterpart. Since this phenomenon can already
be observed for a low number-of-bodies case, let us again start by
presenting an example. 

\begin{example}[$N=2$]
\label{example:discrete-N=2}
    For the two-body problem, the symmetry algebra is generated by: 
    \begin{equation}
        \widetilde \mathcal{S}^{(2)}  
        = 
        \widetilde \mathcal{S}^{(2)}_\text{f.i} 
        = 
        \set{F_{1},F_{2},\K_{2,1}},
        \label{eq:S2def-dsicr}
    \end{equation}
    where $F_1 $ and $F_2 $ remain unchanged under discretization, \cf{}
    \cref{eq:F1F2K21-N=2-cont-explicit}, and the modified constant of motion,
    from~\eqref{eq:Kdiscr-via-flow},  is given by:
    \begin{equation}
             \K_{2,1} 
                     = K_{2,1} + \alpha \sqrt{h} K^{(1)}_{3, 2}
                    =  ( \vb x \cdot \vb p) F_1 
                    - (x_1 +x_2) F_2 
                    + \frac{1}{2} \alpha \sqrt{h} \, 
                    (x_1 - x_2) (p_1 - p_2)  
            \left(  F_1^2 -  F_2 \right). 
    \end{equation}
    As we already observed, the integral of motion $\K_{2,1}$ is of order
    three.  One can now compute the Poisson commutation relations and arrive at
    the following set of relations: 
\begin{equation}
    \pb{F_1}{F_2}=0,
    \quad
    \pb*{F_1}{\K_{2,1}}  = 2 F_2 - F_1^2,
    \quad
    \pb*{F_2}{\K_{2,1}} =  \alpha \sqrt{h} \left(  F_1^2 - 2 F_2 \right) 
	\left(  F_1^2 -  F_2 \right).
    \label{eq:discr-symmetry-alg-N=2}
\end{equation}
    The first two brackets are unaltered with respect to the continuous case,
    see~\eqref{eq:cont-symmetry-alg-N=2-b}, while the third one is now
    non-zero. This is in line with the fact that, as was
    discussed in \cref{sec:intro}, in the discrete-time setting, there is no
    complete analogue of the Hamiltonian, and being in involution with the
    ``Hamiltonian'' no
    longer implies to be an integral of motion, and \vv{}. It
    is easy to see that in the continuum  limit we have: 
\begin{equation}
	\lim_{h \to 0} \{ F_2, \K_{2,1} \} = \{F_2, K_{2,1} \} = 0. 
\end{equation}
    In this sense, the polynomial algebra $\widetilde \mathcal{A}^{(2)} $,
    generated by $ \widetilde \mathcal{S}^{(2)}$ can be seen as a
    \emph{deformation} of the continuous symmetry algebra $ \mathcal{A}^{(2)} $
    with respect to the discretization parameter~$h$. We remark that the following nontrivial phenomenon occurs: since the discretization of $K_{m,n} $
    increases its order, the associated polynomial symmetry algebra becomes of higher degree in the generators. This will be shown to be true below for an
    arbitrary $N$. For $N=2$, the continuous symmetry algebra was quadratic,
    while the corresponding deformed symmetry algebra in the discrete case turns out to be quartic. 
\end{example} 
Let us now proceed to analyze the structure of the symmetry algebra for an arbitrary $N$-body NPdCM model. Taking into account that in the discrete case, the integral of motion $\K_{m,n} $ \eqref{eq:discr-Kmk} can be rewritten via the constant  of motion $K_{m,n} $ of
the continuous model and the term $K_{m+1, n+1}^{(1)} $
(see~\cref{eq:Kdiscr-via-flow}), we can use the results of
\cref{sec:continuous-CM}.

First, directly from \cref{prop:commform} and taking into account the presence of the additional term, one can obtain the following result:
\newpage
\begin{prop}\label{prop:commform-discrete}
    The following formal expressions for the Poisson brackets hold true in the
    discrete case:      
\begin{subequations}\label{eq:commform-discrete}
	\begin{align}
        \pb*{F_i}{\K_{m,n} } = &  \phantom{+}  i (F_m F_{i+n-2} 
	-  F_n F_{i+m-2})  
	  +i  \alpha \sqrt{h} \ \left(
	  F_m F_{i+n-1} 
		 - F_n F_{i+m-1}
		   \right ), 
		   \label{eq:FK-discrete-closed}
     \\
     \pb*{\K_{i,j}}{\K_{m,n} } = & \phantom{+}   
     F_i ( m \K_{n, j+m-2} 
		+ n \K_{j +n - 2, m}  ) 
		+ F_j ( m \K_{i+m-2, n}  
		  +n \K_{m,i+n-2} )   
		  \notag 
		  \\
		  & + F_m ( i \K_{i+n-2, j}+ 
		  j \K_{i, j+n-2} )  
		  + F_n ( i \K_{j,i+m-2}
		    + j \K_{j+m-2,i} )  
		  \notag 
		  \\
		  &+ \alpha \sqrt{h}  \bigl[   
		  F_i ( m \K_{n, j+m-1} 
		+ n \K_{j +n - 1, m}  ) 
		+ F_j ( m \K_{i+m-1, n}  
		  +n \K_{m,i+n-1} )   
		  \notag 
		  \\
		  & \phantom{+ \alpha \sqrt{h}  \bigg[ } 
		    + F_m ( i \K_{i+n-1, j}+ 
		  j \K_{i, j+n-1} ) 
		  + F_n ( i \K_{j,i+m-1}
		    + j \K_{j+m-1,i} )  \bigr] .
		   \label{eq:KK-expressed-discr}
\end{align}
\end{subequations}
\end{prop}

\begin{remark}
    It is easy to see that in the continuum limit $ h \to 0 $, the
    relations~\eqref{eq:commform-discrete} reduce exactly to their continuous
    counterparts~\eqref{eq:commform}, see~\cref{prop:commform}. From this point
    of view, the symmetry algebra of the discrete model can be considered as a
    \emph{deformed polynomial algebra} with respect to the discretization
    parameter $h$, which plays the r\^ole of deformation parameter. The demonstration of this non-trivial phenomenon constitutes the
    main result of the paper. 
\end{remark}

\begin{proof}[\cref{prop:commform-discrete}]
    By \cref{eq:Kdiscr-via-flow}, we have that:
    $\pb*{F_i}{\K_{m,n}} = \pb*{F_i}{K_{m,n}} 
    + \alpha \sqrt{h}  \pb*{F_i}{ K^{(1)}_{m+1, n+1}} $.
    The explicit form of the bracket $ \{ F_i, K_{m,n} \} $ is known
    (see~\cref{eq:FK-formal-final}). The bracket $\{ F_i, K^{(1)}_{m+1,n+1} \} $, in turn, can be computed as follows: 
    \begin{equation}
            \{ F_i, K^{(1)}_{m+1, n+1} \} = 
             i( F_m F_{i+n-1} 
             -  F_n F_{i+m-1}). 
    \end{equation} 
Combining these expressions, one arrives at~\eqref{eq:FK-discrete-closed}. 

  Using again~\eqref{eq:Kdiscr-via-flow}, one can rewrite the Poisson bracket \eqref{eq:KK-expressed-discr} 
  as follows: 
    \begin{equation}
    \begin{split}
        \pb*{\K_{i,j}}{\K_{m,n}}  
        &= \pb*{K_{i,j} + \alpha \sqrt{h} K^{(1)}_{i+1, j+1}}{%
            K_{m,n} + \alpha \sqrt{h} K^{(1)}_{m+1, n+1}} 
                    \\
                & = \pb*{K_{i,j}}{K_{m,n}}
            + \alpha \sqrt{h} \pb*{ K_{i,j}}{K^{(1)}_{m+1, n+1}} 
            + \alpha \sqrt{h} \pb*{ K^{(1)}_{i+1, j+1}}{K_{m,n}} 
        + \alpha^2 h \pb*{ K^{(1)}_{i+1, j+1}}{K^{(1)}_{m+1, n+1}}, 
                \label{eq:KKdiscr-project}
    \end{split}
    \end{equation}
    where we recall that $ K_{i,j} \equiv K_{i,j}^{(2)} $  is an integral of
    motion for the continuous CM model. Hence, the bracket $ \{ K_{i,j},
    K_{m,n} \} $ was already calculated in the continuous case (see
    \cref{eq:them:PB-KK-formal-KK}). The remaining brackets can be computed analogously, using \eqref{eq:genflowa1} and \eqref{eq:commFkJk}.  

    Combining these terms once computed, performing some algebra, and reconstructing the terms
    $\K_{i,j}$ using again~\cref{eq:Kdiscr-via-flow}, one arrives
    at~\eqref{eq:KK-expressed-discr}. 	
\end{proof}

Proceeding as in the continuous setting, we see that for a fixed $N$ to make
the formal Poisson commutation relations~\eqref{eq:commform-discrete} the
actual ones  given in terms of elements in the set $\widetilde \mathcal{S}^{(N)}$, we need to express  the following additional functionally
dependent first integrals:
\newpage
	\begin{subequations}\label{eq:FandKneed-discr}
    \begin{gather}
        \label{eq:FNsneed-discr}
        \begin{array}{ccccc}
            F_{N+1},& F_{N+2},& \ldots, & F_{2N-2}, & F_{2N-1},
        \end{array}
        \\
                \begin{array}{ccccc}
              \K_{1,0}, &   \K_{ 2,0 }, & \ldots, 
              &  \K_{N-1, 0}, &  \K_{N, 0},  
            \end{array}
        \\
        \label{eq:KNspneed-discr}
        \begin{array}{ccccc}
             \K_{N+1,1}, & \K_{N+1,2}, & \ldots, & \K_{N+1,N-1},& \K_{N+1,N},
             \\
             \K_{N+2,1}, & \K_{N+2,2}, & \ldots, & \K_{N+2,N-1},& \K_{N+2,N},
             \\
             \vdots & & & & \vdots 
             \\
             \K_{2N-2,1}, & \K_{2N-2,2}, & \ldots, 
             & \K_{2N-2,N-1},& \K_{2N-2,N},
             \\
             \K_{2N-1,1}, & \K_{2N-1,2}, & \ldots, & \K_{2N-1,N-1} .
         \end{array}
    \end{gather}
    Notice that we need one more row with respect to continuous
    case~\eqref{eq:FandKneed-cont}. 
\end{subequations}

Then, adopting the same procedure shown in the continuous case
(see~\cref{subsec:cont-symmetry-algebra}), one can express the relations,
presented in~\cref{prop:commform-discrete}, in a closed form for a given $N$.
Since the integrals of motion $\set{ F_i }_{i=1}^N $ remain unchanged under
discretization, the same iterative formulas~\eqref{eq:FNstriang} are
applicable. Moreover, by construction the same \cref{eq:KNsNp} holds true upon
the replacement $K_{i,j} \mapsto \K_{i,j}$, \ie{}: 
\begin{equation}\label{eq:Kdiscr-closure}
	 \K_{N+s,\ell}
    =
     -\sum_{k=1}^{N}\frac{1}{k!} 
        B_k \bigl( - 0! F_1, -1! F_2, -2! F_3, \dots, -(k-1)! F_k \bigr)
        \K_{N+s-k,\ell}.
\end{equation}
This allows us to write the explicit form of the symmetry algebra not only for the continuous CM model, but also for its Nijhoff--Pang discretization. We conclude by extending \cref{prop:commrelfin} to the discrete setting:   

\begin{prop}\label{prop:commrelfin-discr}
   The symmetry algebra $\widetilde \mathcal{A}^{(N)}$  of the $N$-body NPdCM model
    is generated by the elements of the set $\widetilde
    \mathcal{S}^{(N)}$~\eqref{eq:SNfidef-discr} with Poisson commutation
    relations given in eq.~\eqref{eq:commform-discrete}, provided the additional closure
    relations given in~\cref{eq:FNstriang,eq:Kdiscr-closure}.
    Moreover, $\mathcal{I}^{(N)} \coloneqq  \langle F_i\rangle_{i=1}^{N}$ is
    an ideal of  $\widetilde{\mathcal{A}}^{(N)}$. Finally, for any $N\geq 2$, the degree of the polynomial symmetry algebra $\widetilde{\mathcal{A}}^{(N)}$ is $2N $.
\end{prop} 

\begin{proof}
As for the continuous case, everything in the previous proposition was already
proved except for the degree of the symmetry algebra. To determine the degree of the
symmetry algebra in the discrete case, we can use the same argument as in the
proof of \cref{prop:commrelfin}, \mm. Hence,  taking into account the new
terms proportional to the discretization parameter in
\cref{eq:commform-discrete}, which have a higher
degree, we obtain the thesis. 
\end{proof}

    We stress that the discrete symmetry algebra $\widetilde{\mathcal{A}}^{(N)}$  reduces to the continuous one 
    in the continuum limit, \ie{} $h \to 0 $, without the necessity of rescaling other
parameters. That is, the following result holds:
\begin{corollary}\label{th:corol-main}
    The symmetry algebra  $\widetilde \mathcal{A}^{(N)}$  of the $N$-body
    NPdCM model is a deformation of the symmetry algebra $ \mathcal{A}^{(N)} $
    $N$-body CM model, that is the Poisson commutation relations of
    $\widetilde \mathcal{A}^{(N)}$ reduce to those of  $ \mathcal{A}^{(N)} $
    by letting the discretization parameter~$h \to 0 $. 
\end{corollary}
The above result relates the discrete and continuous CM systems on an algebraic level and constitutes a remarkable property of the model.

\begin{remark}
    We remark that the discretization of the CM model not only deforms the
    symmetry algebra with respect to the deformation parameter $h$, but also
    increases its degree, provided by the contribution of the new additional terms,
    proportional to $h$. For $N>2$, we have $\deg \widetilde \mathcal{A}^{(N)}=
    \deg \mathcal{A}^{(N)}+1$, \ie{} the degree of the polynomial algebra of
    the CM model is increased by one, while for $N=2$ we have $\deg \widetilde
    \mathcal{A}^{(2)}= \deg \mathcal{A}^{(2)}+2$, \ie{} the degree is raised by
    two. 
    \label{rem:degcontvsdiscr}
\end{remark}

We conclude this section with a detailed study of the discrete
three-body case.
\begin{example}[$N=3$]
    In the same fashion as we have done for the continuous CM model
    (\cf{}~\cref{example:cont-N=3}), let us now demonstrate an application of
    the general formulas~\eqref{eq:commform-discrete} to the three-body NPdCM
    system.  From~\eqref{eq:SNdef-discr}, we have that the symmetry algebra is
    generated by the set: 
    \begin{equation}\label{eq:discr-S3}
       \widetilde \mathcal{S}^{(3)} = \Set{F_1,F_2,F_3,\K_{2,1},\K_{3,1},\K_{3,2}}.
    \end{equation}
    Since the integrals of motion $\set{F_i}_{i=1}^N$ remain unchanged under
    discretization, their explicit expression is given by the same
    formulas~\eqref{eq:constants3DcaseF}. The modified additional constants of
    motion, in turn, can be written as follows:
    \begin{subequations}
    \begin{align}
            \K_{2,1} & = K_{2,1} +  \alpha \sqrt{h} \ \big[ (\vb x \vdot  \vb A) F_1 
               -  (\vb x \vdot \vb p) F_2  \big], 
            \\
            \K_{3,1} & = K_{3,1} +  \alpha \sqrt{h} \ 
            \left[ \   \left( 2  \sum_{i=1}^3 x_i p_i A_i 
                     -  \sum_{i=1}^3 x_i p_i^3 
                     + \nu^2  \left( Y_{1,2} + Y_{1,3} + Y_{2,3} \right) 
                     \right) F_1
                      -  (\vb x \cdot \vb p) F_3 \right],
        \\
        \K_{3,2} &= K_{3,2} +    \alpha \sqrt{h} \ 
            \left[ \ \left( 2  \sum_{i=1}^3 x_i p_i A_i 
                     -  \sum_{i=1}^3 x_i p_i^3 
                     + \nu^2  \left( Y_{1,2} + Y_{1,3} + Y_{2,3} \right) 
                     \right) F_2
                      - (\vb x \cdot \vb A) F_3 \right], 	 
    \end{align}
    \end{subequations}
    where $K_{i,j} $ are integrals of motion for the continuous system
    (see~\cref{eq:constants3DcaseK}),  $\vb A $ is defined again by
    \cref{eq:vecA-def}, and we also introduced the following auxiliary quantity: 
\begin{equation}\label{eq:Phi-def}
	 Y_{i,j}  \coloneqq \frac{p_i x_j + p_j x_i}{(x_i -x_j)^2}.  
\end{equation}
    Let us proceed to compute the Poisson commutation relations of the
    symmetry algebra $\widetilde \mathcal{A}^{(3)}$ for the discrete model,
    generated by the elements in the set $\widetilde \mathcal{S}^{(3)}$. The abelian part of the
    algebra, generated by $\set{F_{1},F_{2},F_{3}}$, is preserved. 
    Then, on a formal level, we can compute the relations,
    using formula~\eqref{eq:FK-discrete-closed}: 
    \begin{subequations}\label{eq:discr-example-N=3-formal-relations}
    \begin{align}
        \pb*{F_1}{\K_{2,1}} &= -F_{1}^2+3F_{2}, 
        \\
        \pb*{F_1}{\K_{3,1}} &= -F_{1}F_{2}+3F_{3},
        \\
        \pb*{F_1}{\K_{3,2}} &=  F_{1} F_{3}-  F_{2}^2,
        \\
        \pb*{F_2}{\K_{2,1}} & = 2 \alpha \sqrt{h} 
        \left(  F_2^2 -  F_1 F_3   \right), 
        \\
        \pb*{F_2}{\K_{3,1}} & =   2\alpha \sqrt{h} 
           \left(  F_2 F_3 - F_1 F_4   \right), 
         \\
        \pb*{F_2}{\K_{3,2}} & = 2\alpha \sqrt{h} \left(  F_3^2 -  F_2 F_4   \right),
        \\
        \pb*{F_3}{\K_{2,1}}  &= 3 (F_{2}^2 - F_{1}F_{3})
        +  3 \alpha \sqrt{h} ( F_2 F_3 - F_1 F_4 ), 
               \\
         \pb*{F_3}{\K_{3,1}}  &=  3(F_{2} F_{3}- F_{1}F_{4} ) 
         +3\alpha \sqrt{h} (F_3^2 - F_1 F_5 ),
          \\
          \pb*{F_3}{\K_{3,2}} & = 3 (F_{3}^2- F_{2}F_{4}) 
          + 3\alpha \sqrt{h} ( F_3 F_4 - F_2 F_5 ),	
    \end{align}	
    and from~\eqref{eq:KK-expressed-discr}: 
    \begin{align}
    \begin{split}
     \pb*{\K_{2,1}}{\K_{3,1}} & = 
       2 F_1 \K_{3,1}  
	       -F_2 ( 3\K_{2,1} + \K_{3,0})
	       + F_3 \K_{2,0}
	       \\ & \phantom{+} 
	       +  \alpha \sqrt{h} \left(  3 F_3 \K_{2,1}  - 4 F_2 \K_{3,1} 
	       +  F_1 ( 2 \K_{3,2} + \K_{4,1} )
     \right),   	
    \end{split} 
                   \\
                     \begin{split}
    \pb*{\K_{2,1}}{\K_{3,2}} & =  
    5 F_{1} \K_{3, 2} 
	       - 4 F_2  \K_{3,1} + 3 F_3 \K_{2,1}  
	       \\ & \phantom{+}   
	       +2\alpha \sqrt{h} \left( \frac{3}{2} F_1 \K_{4,2} 
	       + F_3 \K_{3,1}
    -  F_2 \K_{4,1} - 2 F_2 \K_{3,2} \right),  
      \end{split} 
                   \\
    \begin{split}
         \pb*{\K_{3,1}}{\K_{3,2}} & =  
     3 F_1 \K_{4,2} - F_2 ( \K_{3,2} + 3 \K_{4,1} ) 
	       +  2 F_3 \K_{3,1}  
     \\ & \phantom{+}  + \alpha \sqrt{h} ( 
     3 F_1 \K_{5,2} 
     - 3 F_2 \K_{5,1} 
      + 3 F_3\K_{4,1}
     -  2F_1 \K_{4,3}
     - 4 F_3 \K_{3,2}
     ).  	
     \end{split}	
    \end{align}
    \end{subequations}
Following the same algorithm as in the continuous case, we now have to express the elements $F_4,F_5, \K_{2,0}, \K_{3,0}, \K_{4,1}, \K_{4,2}, \K_{4,3}, \K_{5,1}, \K_{5,2}$ in terms of the generators~\eqref{eq:discr-S3}.  We have, similarly to the continuous case (see~\eqref{eq:funcrel-discr} and \eqref{eq:funcrel-K0}) that $  F_3 \K_{2,0} - F_2 \K_{3,0} = - 3 \K_{3,2} $.   Then, from~\eqref{eq:FNs} and~\eqref{eq:Kdiscr-closure}, we obtain: 
\begin{subequations}\label{eq:F4F5K41K42K43K51K51-discr}
\begin{align}
 F_4 &= \frac{4}{3} F_{1}F_{3} - F_{1}^2F_{2} +\frac{1}{2}F_{2}^2 
            +\frac{1}{6}F_{1}^4 ,
 \\
 F_5 &= \frac{5}{6} ( F_1^2 F_3 -  F_1^3 F_2 +  F_2 F_3)
 + \frac{1}{6} F_1^5  , 
 \\
\K_{4,1} &= F_{1} \K_{3, 1}
            - \frac{1}{2} \left(  F_{1}^2- F_{2}\right) \K_{2, 1},
\\
\K_{4,2} &= F_{1} \K_{3, 2} 
            - \frac{1}{6} \left( F_{1}^3 - 3 F_{1}F_{2} + 2 F_{3}\right) \K_{2, 1},
\\
\K_{4,3} & =  - \frac{1}{6} \left( F_{1}^3 -3 F_{1}F_{2}+2F_{3}\right) \K_{3,1} + \frac{1}{2} \left(  F_1^2 -F_2 \right) \K_{3,2},
\\
\K_{5,1} & = \frac{1}{3} \left(F_3 - F_1^3 \right)\K_{2,1}
 + \frac{1}{2} \left( F_1^2 +F_2 \right) \K_{3,1},
\\
\K_{5,2} & =     
\left( \frac{1}{2} F_1 F_2 - \frac{1}{3} F_3 - \frac{1}{6} F_1^3 \right) F_1 \K_{2,1}
+ \frac{1}{2}  \left( F_1^2 +F_2 \right) \K_{3,2}.         
\end{align}	
\end{subequations}
Combining~\eqref{eq:F4F5K41K42K43K51K51-discr} with~\eqref{eq:discr-example-N=3-formal-relations}, we finally arrive at the closed-form expressions for the Poisson commutation relations, completely defining the structure of the symmetry algebra $\widetilde \mathcal{A}^{(3)}$: 
\begin{subequations}
\begin{align}
	\pb*{F_1}{\K_{2,1}} &= -F_{1}^2+3F_{2}, 
	\label{eq:discrN=3:F1K21}
	\\
	\pb*{F_1}{\K_{3,1}} &= -F_{1}F_{2}+3F_{3},
	\label{eq:discrN=3:F1K31}
	\\
	\pb*{F_1}{\K_{3,2}} &=  F_{1} F_{3}-  F_{2}^2,
	\label{eq:discrN=3:F1K32}
	\\
	\pb*{F_2}{\K_{2,1}} & =  2 \alpha \sqrt{h} 
	\left( F_2^2 -  F_1 F_3   \right), 
	\label{eq:discrN=3:F2K21}
	\\
	\pb*{F_2}{\K_{3,1}} & = \frac{1}{3}  \alpha \sqrt{h} 
	   \left[ 6 F_2 F_3 -  F_1 
	   \left( 8 F_{1}F_{3} -6F_{1}^2F_{2} +3 F_{2}^2  + F_{1}^4 \right)   \right], 
	\label{eq:discrN=3:F2K31} 
	\\
	\pb*{F_2}{\K_{3,2}} & = \frac{1}{3} \alpha \sqrt{h} \left[ 
	      6 F_3^2 -  F_2 
	\left( 8 F_{1}F_{3} - 6 F_{1}^2F_{2} 
	+3F_{2}^2  +F_{1}^4 \right)   \right],
	\label{eq:discrN=3:F2K32}
	\\
	\pb*{F_3}{\K_{2,1}}  &=  3(F_{2}^2 -F_{1}F_{3} )	
	+ \frac{1}{2}  \alpha \sqrt{h} \left[ 6 F_2 F_3 
	- F_1 
	 \left( 8F_{1}F_{3} - 6 F_{1}^2F_{2} 
	 +3 F_{2}^2  + F_{1}^4 \right) \right],
	 \label{eq:discrN=3:F3K21}  %
	       \\ 
	  \begin{split}
	 \pb*{F_3}{\K_{3,1}}  &=     3F_{2} F_{3}-4F_{1}^2F_{3} +3F_{1}^3 F_{2} - \frac{3}{2} F_{1}F_{2}^2 
            - \frac{1}{2}F_{1}^5
            \\ & \phantom{+} 
            + \frac{1}{2} \alpha \sqrt{h} 
            \left(  F_3 - F_1^3 \right)
            \left(  F_1^3 -5 F_1 F_2 + 6 F_3 \right), 
            \label{eq:discrN=3:F3K31}
            	  \end{split} 
              \\
    \begin{split}
        \pb*{F_3}{\K_{3,2}} &=    3F_{3}^2-4F_{1}F_{2}F_{3} +3 F_{1}^2 F_{2}^{2} 
               -\frac{3}{2} F_{2}^3 
            -\frac{1}{2}F_{1}^4F_{2}
               \\
          & \quad  + \frac{1}{2} \alpha \sqrt{h} 
             \left( F_1^4 F_3 
            - F_1^5 F_2
                     + 5 F_1^3 F_2^2  
            - 11 F_1^2 F_2 F_3 
            + 8 F_1 F_3^2
            -2 F_2^2 F_3
            \right),   
            \label{eq:discrN=3:F3K32} 
     \end{split}  
     \\
     \begin{split}
      \pb*{\K_{2,1}}{\K_{3,1}}  &=  2F_{1}\K_{3, 1}-3F_{2}\K_{2, 1}-3\K_{3, 2}    
           \\
     & \quad  + \frac{1}{2} \alpha \sqrt{h} \left[ 
        \left( F_1 F_2 + 6 F_3 - F_1^3 \right) \K_{2,1}
      +2 ( F_1^2 - 4 F_2 ) \K_{3,1} 
        + 4 F_1 \K_{3,2} 
        \right],  
        \label{eq:discrN=3:K21K31}
     \end{split} 
     \\
     \begin{split}
     	   \pb*{\K_{2,1}}{\K_{3,2}}  &=  5F_{1}\K_{3, 2} -4F_{2}\K_{3,1} 
           +3F_{3}\K_{2, 1}
         \\
         & \quad  + \frac{1}{2} \alpha \sqrt{h} \left[
         \left( 5 F_1^2 F_2 - 2 F_1 F_3 - 2 F_2^2   -  F_1^4 \right) 
         \K_{2,1} 
         +4 ( F_3 -  F_1 F_2 ) \K_{3,1}
         +2 (3 F_1^2 - 4F_2) \K_{3,2}
         \right],  
         \label{eq:discrN=3:K21K32}
     \end{split}
     \\
     \begin{split}
      \pb*{\K_{3,1}}{\K_{3,2}}  &=   2F_3\K_{3, 1}
	       	 +3F_1^2 \K_{3, 2} 
	       	 -\frac{1}{2}F_{1}^4 \K_{2, 1}
           +3 F_{1}^2F_{2} \K_{2,1} - F_{1}F_{3}\K_{2, 1}
           -3F_{1}F_{2} \K_{3, 1} 
           \\
           & \quad  
           -\frac{3}{2} F_{2}^2 \K_{2, 1}-F_{2} \K_{3, 2}
           + \frac{1}{2} \alpha \sqrt{h} \left[ 
           \left( 5 F_1^3 F_2 +  F_2 F_3 
           - 5 F_1^2 F_3 -  F_1^5  \right) \K_{2,1} 
           \right.
           \\
           & \quad 
           \left. 
          + \left(   \frac{2}{3} F_1^4 -5 F_1^2 F_2 
           + \frac{22}{3} F_1 F_3 -  3 F_2^2  \right) \K_{3,1}
           + \left(  F_1^3  + 5 F_1 F_2 -  8 F_3  \right) \K_{3,2} 
           \right].
           \label{eq:discrN=3:K31K32}
     \end{split}
\end{align}	
\end{subequations} 
The functional relation is given by: 
\begin{equation}
	 F_2 \K_{3,1} - F_1 \K_{3,2} 
	- F_3 \K_{2,1} =0 
\end{equation}
Let us compare this algebra with the continuous case (see~\cref{example:cont-N=3}).
Note that the relations~\eqref{eq:discrN=3:F1K21}-\eqref{eq:discrN=3:F1K32} are preserved, \ie{} they coincide with the continuous counterparts~\eqref{eq:F1K21N3}-\eqref{eq:F1K32N3}.   

The three-body example reveals the phenomena in more detail, compared to~\cref{example:discrete-N=2}.   For instance, here not only the relations $\{F_2, \hyphen \} $ become non-zero (as it was already discussed before), but also the relations~\eqref{eq:discrN=3:F3K21}-\eqref{eq:discrN=3:K31K32}, being already nontrivial in the continuous case (\cf{}~\eqref{eq:F3K21N3}-\eqref{eq:K31K32N3}), are now deformed with respect to $h$ as well. This clearly follows from the general formulas~\eqref{eq:commform-discrete}.

\end{example}

\section{Discussion and outlook}
\label{sec:conclusions}

In this paper, we computed the explicit symmetry algebras for the celebrated
 $N$-body Calogero--Moser model and its discretization, introduced by Nijhoff and Pang,
for an arbitrary number of bodies~$N$. The structure of the symmetry algebra is
given by its Poisson commutation relations and closure relations in the form of
recursive formulas (see~\cref{prop:commrelfin} and \cref{prop:commrelfin-discr}
for the continuous and discrete models, respectively). The final formulas,
obtained through a judicious application of Cayley--Hamilton theorem, contain a
peculiar class of special polynomials: the Bell polynomials. We underline that
Bell polynomials should appear in every case, either continuous or discrete, 
when the integrals of motion are given by traces and superintegrability is obtained through an additional isospectral problem
like~\eqref{eq:eom-forX-cont}. Finally, we illustrated the results by
presenting the low-number-of-bodies cases $N=2,3$.

Comparing the obtained symmetry algebras, we demonstrated a remarkable
property: the symmetry algebra of the discrete case is a nontrivial
deformation of the one obtained in the continuous case with respect to the
discretization parameter $h$. Moreover, this deformation does not preserve
the degree of the continuous symmetry algebra: in the discrete setting, higher degree terms emerge, increasing the degree of the polynomial
algebra.  To be more precise, for $N>2$, the degree increases by one, from
$2N-1$ to $2N$, whereas for $N=2$ it increases by two, from two to four.  In the case of the CM symmetry algebra, the contraction from
the deformed to the non-deformed algebra is simply given by taking the
continuous limit $h \to 0 $. To our knowledge, this is the first example of
such a phenomenon. Therefore, our central result is that maximally superintegrable discretizations can naturally yield deformed polynomial Poisson algebras. We observe that the deformation of the symmetry algebra we exhibit in this paper is possible because of the presence of modified integrals of motion of higher order for the discrete system. This is in contrast with the much simpler example of the harmonic oscillator we showed
in~\cite{DGLExplicitIsomorphisms2025}, where the discretized invariants are
modified, but of the same order as the continuous ones, and the same holds true
for the corresponding symmetry algebras. For this reason, in this case, a condition for a non-trivial deformation to happen is that the discrete invariants are modified in a nontrivial way, \eg{} with their order being increased. We observe that,
in the example we considered in this paper, and as well in the one we considered
earlier in~\cite{DGLExplicitIsomorphisms2025}, the deformation in the discretization, whether trivial or non-trivial, appears in the integrals of motion, while the Poisson bracket is kept canonical. Following the celebrated
Ge--Marsden theorem~\cite{ZhongMarsden1988}, different discretizations can
arise by keeping the integrals of motion as in the continuous counterpart and
deforming the Poisson bracket. Clearly, a final possibility would be to
discretize deforming \emph{both} the Poisson bracket and the integrals of
motion.

Our findings potentially suggest an inverse problem: can one find a new discretization of a given continuous model, preserving the superintegrable properties, by deforming a known symmetry algebra? This points a new direction of research and expands the potential for applying the algebraic approach in search of new discretizations. 
We reserve to address the problem of carrying out a systematic search for
discrete MS models whose invariants and/or Poisson
bracket are non-trivially modified for future research. Here, we limit ourselves
to comment that, since a similar result is known for the continuous Toda
lattice~\cite{Agrotisetal2006}, a natural system to check will be its discretization, see \eg~\cite[\S 3.10]{surisProblemIntegrableDiscretization2003}.
Another natural candidate to consider is the CM system with an external
harmonic potential or Kepler--Coulomb potential (see, for example, \cite{PhysRevD.93.125008} and references therein), whose discretization was
considered in~\cite{Ujino_etal2005}. To our knowledge, the symmetry algebra of
the (discrete) CM model with an external potential has not been determined
yet, and due to the differences between this model and the ``free'' one, in our
opinion, it poses already a challenging problem. Finally, another possible direction could be an application of the obtained results in the context of modern representation theory and Calogero--Moser spaces~\cite{etingofCalogeroMoserSystemsRepresentation2007}.

\section*{Acknowledgements}

PD, GG, and DL  acknowledge the support of the research project Mathematical
Methods in NonLinear Physics (MMNLP), Gruppo 4-Fisica Teorica of INFN.  This
work has been partially supported by the  National Group of Mathematical
Physics (GNFM) of the Italian Institute for High Mathematics (INdAM).  

PD acknowledges the support of the Universit\`a degli Studi di Milano (for partially funding the research visit during which significant progress on this work was made), and of the~Ph.D.~program of the
Universit\`a degli Studi di Udine.  

DL has been partially funded by MUR - Dipartimento di Eccellenza 2023-2027,
codice CUP G$43C22004580005$ - codice progetto   DECC$23$$\_$$012$$\_$DIP.

\appendix

\section{Symmetric polynomials and Newton's identity}
\crefalias{section}{appendix}

\label{app:symmpoly}

In this appendix, for the convenience of the reader, we report some results on the
symmetric polynomials and Newton's identity that we need in the text. For a more
general presentation of the topic, we refer
to~\cite{macdonaldSymmetricFunctionsHall1995} and references therein.

Let $\vb{x} \coloneqq ( x_1, \dots, x_N ) \in \R^N $.  Then, the
\emph{$k$\textit{-th} power sum} is:
\begin{equation}\label{eq:app-power-sums}
	f_k (\vb{x}) \coloneqq \sum_{i=1}^N x_i^k = x_1^k + \cdots + x_N^k. 
\end{equation}
Moreover, the \emph{elementary symmetric polynomial} $e_k (\vb  x)$ is defined
to be the sum of all distinct products of $1\leq k\leq N$ variables $\{x_1,
\dots, x_N\}$:
\begin{equation}
    e_{k}(\vb{x}) \coloneqq  
    \sum_{i_{1}<\ldots<i_{k}} 
    x_{i_{1}}\cdots x_{i_{k}}.
    \label{eq:symmpoly}
\end{equation}
Symmetric polynomials appear naturally in the factorization of a
single-variable monic polynomial of degree $N$ with roots
$x_{1},\ldots,x_{N}$~\cite{tignolGaloisTheoryAlgebraic2016}:
\begin{equation}
    \prod_{i=1}^{N}(X-x_{i})
    =
    X^N - e_1(\vb{x}) X^{N-1} + e_2(\vb{x}) X^{N-2} 
    - e_3(\vb{x}) X^{N-3} + \cdots + (-1)^N e_N(\vb{x}) =0.
    \label{eq:factopoly}
\end{equation}

The relation between elementary symmetric polynomials and power sums
is given by the following result:

\begin{newton}
 The following identity holds true: 
 \begin{equation}\label{eq:Newtons-identity}
     e_k(\vb{x}) = \frac{1}{k} \sum_{i=1}^k (-1)^{i-1} e_{k-i}(\vb{x}) f_i(\vb{x}). 
 \end{equation}
\end{newton}

From Newton's identity~\eqref{eq:Newtons-identity}, it is possible to express the elementary symmetric polynomials in terms of the power sums: 
\begin{equation}\label{eq:McDnld-f-la-det}
	e_k(\vb x) = \frac{1}{k!} 
	\det \begin{bmatrix}
		f_1 (\vb x) & 1 & 0 & \cdots &0
		\\
		f_2 (\vb x)  & f_1 (\vb x) & 2  &  & 0
		\\
		\vdots & \vdots & \ddots & \ddots   & \vdots 
		\\
		f_{k-1} (\vb x) &  f_{k-2}(\vb x) &  & \ddots  & k-1
		\\
		f_k (\vb x)  & f_{k-1} (\vb x) & \cdots & \cdots  & f_1 (\vb x) 
	\end{bmatrix}. 
\end{equation}
Lastly, we observe that equation~\eqref{eq:McDnld-f-la-det} can be expressed
through the so-called \emph{complete exponential Bell polynomials}, an infinite
family of inhomogeneous polynomials of total degree $k$ in $k$ variables,
explicitly given by~\cite[\S 3.3]{comtetAdvancedCombinatoricsArt1974}:
\begin{equation}
    B_{k}(y_{1},\ldots,y_{k}) \coloneqq  
    k! \sum_{1j_{1}+\ldots+kj_{k}=k} 
    \prod_{i=1}^{k} \frac{y_{i}^{j_{i}}}{(i!)^{j_{i}}j_{i}!}.
    \label{eq:bell}
\end{equation}
Indeed, through the properties of determinants, it is possible to see that
we can rewrite formula~\eqref{eq:McDnld-f-la-det} through~\eqref{eq:bell} 
as:
\begin{equation}\label{eq:sym-bell}
	e_k(\vb x) = \frac{(-1)^k}{k!} B_k 
	\bigl( -f_1(\vb x), \  -1! f_2(\vb x), \  -2! f_3(\vb x), \ \dots, \ 
	-(k-1)! f_k(\vb x)  \ \bigr). 
\end{equation}
Let us now recall that given a $N\times N$ matrix $A$,  its trace is a sum of its
eigenvalues:
\begin{equation}
    \trace A = \lambda_1 + \cdots + \lambda_N. 
\end{equation}
This readily implies that, for all $k\in \N$, we have the identity:
\begin{equation}\label{eq:app-trAk-power-sum}
\trace (A^k) = \sum_{i=1}^N \lambda_i^k= 
	f_k  (\vb*  \lambda),
\end{equation}
where we used the power sum notation~\eqref{eq:app-power-sums}.

Next, let us
apply the Cayley--Hamilton theorem~\cite{loehrAdvancedLinearAlgebra2024} to the matrix $A$. This famous result states that a matrix solves its own
characteristic equation:
\begin{equation}\label{eq:matrix-pol-CHthm}
	A^N -e_1 (\vb*{\lambda} ) A^{N-1} 
	    + e_2 (\vb*{\lambda} ) A^{N-2}
	  + \cdots + (-1)^{N-1}e_{N-1} (\vb*{\lambda} ) A + (-1)^{N}e_{N} (\vb*{\lambda} ) \1_N = \mathbb{0}_N,
\end{equation}
where  $\mathbb{0}_N$ and $\1_N $ are zero and identity matrices of size
$N \times N$ respectively, and we used~\eqref{eq:factopoly} to express the
coefficient of the characteristic polynomial in terms of its roots, namely the
eigenvalues $\vb*{\lambda}=(\lambda_1,\ldots,\lambda_N)$.  Now, multiplying the
relation~\eqref{eq:matrix-pol-CHthm} by the matrix~$A^s$ for $s\in\N$ and
taking the traces, using~\eqref{eq:app-trAk-power-sum}, we obtain:
\begin{equation}\label{eq:FNstraces}
	f_{N+s} =
    \sum_{i=1}^N (-1)^{i+1} e_i (\vb* \lambda) f_{N+s-i}.
\end{equation}    
Therefore, using~\eqref{eq:sym-bell}, we arrive at:
\begin{equation}\label{eq:app-arbitrary-matr-traces-Bell}
	f_{N+s} = - \sum_{k=1}^N \frac{1}{k!} 
	B_k \bigl( -0! f_1, -1! f_2, -2! f_3, \ldots,  
	- (k-1)! f_k \bigr) f_{N+s-k},
\end{equation}
for $s>0 $.  The formula~\eqref{eq:app-arbitrary-matr-traces-Bell} allows us to  express 
$f_{N+s}$ in terms of $f_{1},\ldots,f_{N+s-1}$. Thus, by applying this formula recursively, one can express $f_{N+s} $ in terms of~$ \set{f_1, \dots, f_N} $, see also~\cite{Lavoie1975}.   

\footnotesize
\bibliographystyle{alphaurl}

\bibliography{bib}

\end{document}